\documentclass[12pt]{article}
\evensidemargin=\oddsidemargin
\usepackage{color}
\usepackage{cite}

\usepackage{hyperref}

\definecolor{Red}           {cmyk}{0,1,1,0}
\definecolor{Black}         {cmyk}{0,0,0,1}

\newcounter{bla}
\newenvironment{refnummer}{%
\list{[\arabic{bla}]}%
{\usecounter{bla}%
 \setlength{\itemindent}{0pt}%
 \setlength{\topsep}{0pt}%
 \setlength{\itemsep}{0pt}%
 \setlength{\labelsep}{2pt}%
 \setlength{\listparindent}{0pt}%
 \settowidth{\labelwidth}{[9pt]}%
 \setlength{\leftmargin}{\labelwidth}%
 \addtolength{\leftmargin}{\labelsep}%
 \setlength{\rightmargin}{0pt}}}
 {\endlist}

\newcommand{\imag}{\Im {\rm m}}
\newcommand{\real}{\Re {\rm e}}

\newcommand{\ra}{\rightarrow}
\newcommand{\bea}{\begin{eqnarray}}
\newcommand{\eea}{\end{eqnarray}}

\newcommand{\bra}[1]{\langle #1|}
\newcommand{\ket}[1]{|#1\rangle}

\newcommand{\bc}{\begin{center}}
\newcommand{\ec}{\end{center}}

\def\beq{\begin{equation}}
\def\eeq{\end{equation}}
\def\beas{\begin{eqnarray*}}
\def\eeas{\end{eqnarray*}}
\def\ba{\begin{array}}
\def\ea{\end{array}}

\def\code{{\tt SUSY\_FLAVOR}}

\def\webpage{{\tt http://www.fuw.edu.pl/susy\_flavor}}

\begin{document}



\thispagestyle{empty}

\begin{flushright}
September 20, 2014
\end{flushright}
\vspace{-3mm}
\begin{center}
{\Large \bf {\color{Red}\code~}{\color{Black} v2.5: a computational
    tool for FCNC and \\[3mm] CP-violating processes in the MSSM}}
\\[0.8cm]

{\large A.~Crivellin$^a$, J.~Rosiek$^b$, P.~Chankowski$^b$,
  A.~Dedes$^c$, S.~J\"ager$^d$
  P.~Tanedo$^e$}\\[0.5cm]

  {\em $^a$Albert Einstein Center for Fundamental Physics, Institute
    for Theoretical Physics, University of Bern, CH-3012 Bern,
    Switzerland}\\[0.2cm]

  {\em $^b$Institute of Theoretical Physics, University of Warsaw,
    00-681 Warsaw, Poland}\\[0.2cm]

  {\em $^c$Division of Theoretical Physics, University of Ioannina, GR
    45110, Greece}\\[0.2cm]

  {\em $^d$ Department of Physics and Astronomy, University of Sussex,
     Brighton BN1 9QH, UK}\\[0.2cm]

  {\em $^e$Institute for High Energy Phenomenology, Newman Laboratory
    of Elementary Particle Physics, Cornell University, Ithaca, NY
    14853, USA}

\vskip 3mm

Abstract\\[2mm]
\begin{minipage}[t]{14.5cm}
\small
\noindent We present \code{} version 2.5 --- a Fortran 77 program that
calculates low-energy flavor observables in the general $R$-parity
conserving MSSM.  For a set of MSSM parameters as input, the code
gives predictions for:\\
$\phantom{~~~~}$ 1. Electric dipole moments of the leptons and the
neutron.  \\
$\phantom{~~~~}$ 2. Anomalous magnetic moments (i.e.  $g-2$) of the
leptons.  \\
$\phantom{~~~~}$ 3. Radiative lepton decays ($\mu\to e\gamma$ and
$\tau\to \mu\gamma, e\gamma$).  \\
$\phantom{~~~~}$ 4. Rare Kaon decays ($K^0_L\ra \pi^0\bar\nu\nu$ and
$K^+\ra \pi^+ \bar\nu\nu$).  \\
$\phantom{~~~~}$ 5. Leptonic $B$ decays ($B_{s,d}\ra l^+ l^-$, $B\to
\tau \nu$, $B\to D \tau \nu$ and  $B\to D^\star \tau \nu$).  \\
$\phantom{~~~~}$ 6. Radiative $B$ decays ($B\to\bar X_s \gamma$).\\
$\phantom{~~~~}$ 7. Rare decays of top quark to Higgs boson ($t\to
ch,uh$).\\
$\phantom{~~~~}$ 8. $\Delta F=2$ processes ($\bar K^0$--$K^0$, $\bar
D$--$D$, $\bar B_d$--$B_d$ and $\bar B_s$--$B_s$ mixing).\\
\code{} v2 performs the resummation of all chirally enhanced
corrections, i.e. takes into account the effects enhanced by
$\tan\beta$ and/or large trilinear soft mixing terms to all orders in
perturbation theory.  All calculations are done using exact
diagonalization of the sfermion mass matrices.  Comparing to previous
versions, in \code{} v2.5 parameter initialization in SLHA2 format has
been significantly generalized and simplified, so that program accepts
without modifications most of the output files produced by other codes
calculating MSSM spectra and processes. In addition, the routine
calculating branching ratios for rare decays of top quark to Higgs
boson has been included.  The program can be obtained from \webpage.
\end{minipage}\\[2mm]
\end{center}

\newpage
\setcounter{page}{1}
\tableofcontents
\newpage

\section{Introduction}
\label{sec:intro}

Flavor physics was in the recent years one of the most active and
fastest developing fields in the high energy physics.  Numerous new
experiments, spanning a wide energy range from neutrino mass
measurements to hard proton scattering at LHC collider, helped to
improve significantly the accuracy of various measurements related to
flavor-observables.  Almost all such experiments reported result which
are in agreement with the Standard Model (SM) predictions, with a few
exception where small observed deviations still require further
confirmation (like e.g.  $g-2$ muon magnetic moment
anomaly~\cite{gm2epx,g_2mu}).

The extensive set of measurements available for rare decays puts
strong constraints on the flavor structure of physics beyond the
Standard Model.  In particular, it imposes stringent limits on the
flavor- and CP- violating parameters of the Minimal Supersymmetric
Standard Model (MSSM)~\cite{reviewsMSSM}, where the flavor changing
neutral currents (FCNCs) originate, in addition to the CKM induced
FCNCs, from the fact that one cannot (in general) simultaneously
diagonalize the mass matrices of fermions and sfermions.  Such a
misalignment leads to FCNCs which can involve the strong coupling
constant and which do not necessarily respect the hierarchy of the CKM
matrix.  Moreover, many of the MSSM parameters can take complex values
and are potential sources of CP violation.  Thus supersymmetric
contributions to flavor and/or CP-violating processes can, in
principle, exceed the SM predictions by orders of magnitude.  The
apparent absence of such big effects leads to the constraint that MSSM
couplings which may generate FCNCs and CP violation are actually
strongly suppressed.  The difficulty to explain this suppression is
known as the ``SUSY flavor problem'' and the ``SUSY CP problem''.
Even if one assumes the so-called Minimal Flavor Violation (MFV)
hypothesis \cite{MFV} which requires that {\it all} FCNC effects
originate from the Yukawa couplings of the superpotential,
supersymmetric contributions to various flavor and CP-violating
amplitudes can still be of comparable (or sometimes even much larger,
like in the case of the electron and neutron EDMs or
$B_s\to\mu^+\mu^-$) size as the corresponding SM contributions.

As the accuracy of the flavor experiments constantly improves, it is
important to have an universal computational tool which helps to
compare new data with the predictions of the MSSM.  Developing such a
tool is a non-trivial task requiring extensive and often tedious
calculations.  Numerous analyses have been published in the
literature, but because of the complexity of the problem, they usually
consider only a few rare decays simultaneously.  Furthermore, many
analyses done for general flavor violation in the MSSM use the mass
insertion approximation for the soft terms (MIA) (see e.g.~\cite{MIA,
  MIPORO}) which simplifies the calculations but does not produce
correct results if flavor violation (and/or chirality violation) in
the sfermion sector becomes large.

In a series of papers published since 1997~\cite{MIPORO, POROSA,
  ROS99, BCRS0, ROS01, BCRS1, CHRO, BCRS, BEJR, DRT, ROS09}, many
supersymmetric FCNC and CP-violating observables were analyzed within
the setup of the most general $R$-parity conserving MSSM using exact
diagonalization of the sfermion mass matrices.  A FORTRAN computer
programs based on the common set of Feynman rules of Ref.~\cite{PRD41}
were developed for each process (using also parts of code written for
earlier papers on MSSM Higgs physics~\cite{JRHIGGS}) and, after
collecting them together, published as \code{} v1~\cite{SFLAV,
  SFLAVT}.

\code{} v1 was able to calculate only the 1-loop supersymmetric
virtual corrections, whereas, as widely discussed in the literature
\cite{Hall:1993gn, Carena:1994bv, Hamzaoui:1998nu, Carena:1999py,
  Babu:1999hn, Isidori:2001fv, Isidori:2002qe, DP, BCRS,
  Foster:2006ze, Crivellin:2008mq, Hofer:2009xb, Crivellin:2009ar,
  Girrbach:2009uy, Crivellin:2010gw, Crivellin:2010er}, in the regime
of large $\tan\beta$ or large trilinear $A$-terms so-called chirally
enhanced corrections must be taken into account.  Chiral enhancement
is always related to fermion-Higgs couplings.  Because these couplings
have mass-dimension 4, the corresponding corrections do not vanish in
the decoupling limit ($m_{\rm{SUSY}}\to\infty$) but rather converge to
a constant.  This in turn also means that the flavor-changing neutral
Higgs couplings, which are induced by chirally-enhanced SUSY
corrections, are still relevant for heavy SUSY particles.  Thus,
especially for observables which are sensitive to Higgs contributions
(like for example $B_{s,d}\to\mu^+\mu^-$ or $B_{d,s}$ mixing), the
consistent resummation and inclusion of chirally enhanced corrections
to all orders of perturbation theory is very important.

In Ref.~\cite{JRCRIV} such resummation was performed in the most
general MSSM taking into account all possible sources of chiral
enhancement; in the decoupling limit ($m_{\rm{SUSY}}\gg v_1,v_2$)
analytical formulae has been given. Results of ref.~\cite{JRCRIV} has
been implemented in \code{} v2.0.  The consistent treatment of all
chirally enhanced effects in the general MSSM (and the corresponding
threshold corrections), including correct calculation of neutral Higgs
penguins in such scenario, is a unique feature of \code{} v2 not
shared at the moment by other publicly available programs calculating
rare decays in the supersymmetric models. 

In \code{} v2.5 the input and output routines for reading and writing
files in SLHA2~\cite{SLHA2} format have been significantly generalized
and simplified, so that the program accepts most of the output files
produced by other codes calculating MSSM spectra and processes. The
output of \code{} itself is now by default written to the file {\tt
  susy\_flavor.out} and, as described in more details in
Section~\ref{sec:output} and Appendix~\ref{app:output}, has a
SLHA2-like block structure. \code{} v2.5 allows also for easier
comparison the relative importance of contributions from various MSSM
sectors, providing new debug control variables which allow to
separately switch on/off contributions from diagrams with gauge+Higgs
bosons, gluinos, charginos and neutralinos circulating in loops (see
Section~\ref{sec:debug}). Also, new routines calculating rates of rare
decays of top quark to Higgs boson in the MSSM, based on
Ref.~\cite{T2UH}, has been added.

Several other programs allowing to analyze various aspects of the MSSM
flavor phenomenology have been published.  The most relevant to
\code~are: {\tt CPsuperH}~\cite{Apostolos}, {\tt
  SusyBSG}~\cite{Pietro}, {\tt SPheno}~\cite{Porod}, {\tt
  SuperIso}~\cite{Mahmoudi} and {\tt SUSEFLAV}~\cite{DeSuVe}.  {\tt
  SusyBSG} is dedicated to high-precision predictions for $B\to X_s
\gamma$ while {\tt CPsuperH} and {\tt SuperIso} calculate processes
similar to the ones computed by \code.  However, these existing codes
are restricted to the Minimal Flavor Violation scenario, whereas
\code{} can simultaneously calculate the set of rare decays listed in
Table~\ref{tab:proc} without any (apart from the $R$-parity
conservation) restrictions on the choice of MSSM parameters.
Other publicly available codes that are relevant to \code{} (which can
e.g. calculate the MSSM soft parameters used as input to \code, or for
the same set of input parameters calculate non-FCNC related
observables) are {\tt FeynHiggs}\cite{Sven}, {\tt SoftSUSY}\cite{Ben},
{\tt SuSpect}\cite{Suspect}, {\tt MicrOMEGAs}\cite{micromega}, {\tt
  DarkSUSY}\cite{darksusy} and {\tt NMHDECAY}\cite{nmhdecay}.

\begin{table}[htbp]
\begin{center}
\begin{tabular}{|lcr|}
\hline
Observable & &Experiment \\ \hline

\multicolumn{3}{|c|}{$\Delta F=0$}\\ \hline 

$\frac{1}{2}(g-2)_e$ & &$(1 159 652 188.4 \pm4.3) \times 10^{-12}$~\cite{g_2e_exp} \\

$\frac{1}{2}(g-2)_\mu$ & &$(11659208.7\pm8.7)\times10^{-10}$~\cite{g_2mu} \\

$\frac{1}{2}(g-2)_\tau$ & &$<1.1\times 10^{-3}$~\cite{g_2tau} \\

$|d_{e}|$(ecm) & &$<1.6 \times 10^{-27}$~\cite{de} \\

$|d_{\mu}|$(ecm) & &$<2.8\times 10^{-19}$~\cite{dmu} \\

$|d_{\tau}|$(ecm) & &$<1.1\times 10^{-17}$~\cite{pdg} \\

$|d_{n}|$(ecm) & &$<2.9 \times 10^{-26}$~\cite{dn} \\ \hline

\multicolumn{3}{|c|}{$\Delta F=1$}\\ \hline 

$\mathrm{Br}(\mu\to e \gamma)$ & & $<5.7 \times 10^{-13}$~\cite{mueg}\\ 

$\mathrm{Br}(\tau\to e \gamma)$ & & $<3.3\times 10^{-8}$~\cite{taueg_taumug}\\ 

$\mathrm{Br}(\tau\to \mu \gamma)$ & & $<4.4\times 10^{-8}$~\cite{taueg_taumug}\\ 

$\mathrm{Br}(K_{L }\to \pi^{0} \nu \nu)$ & & $< 6.7\times
10^{-8}$~\cite{Kpinunu} \\

$\mathrm{Br}(K^{+}\to \pi^{+} \nu \nu)$ & &
$17.3^{+11.5}_{-10.5}\times 10^{-11}$~\cite{Kpipnunu} \\

$\mathrm{Br}(B_{d}\to e e)$ & & $<1.13\times 10^{-7}$~\cite{bdee}\\

$\mathrm{Br}(B_{d}\to \mu \mu)$ & & $<7.4\times 10^{-10}$~\cite{bmumu}
\\

$\mathrm{Br}(B_{d}\to \tau \tau)$ & & $<4.1\times
10^{-3}$~\cite{bdtautau} \\

$\mathrm{Br}(B_{d}\to \mu e)$ & & $<3.7\times 10^{-9}$~\cite{bmue}\\

$\mathrm{Br}(B_{s}\to e e)$ & & $<7.0\times 10^{-5}$~\cite{bsee}\\

$\mathrm{Br}(B_{s}\to \mu \mu)$ & & $(2.9\pm 0.7)\times 10^{-9}$~\cite{bsmumu}
\\

$\mathrm{Br}(B_{s}\to \tau \tau)$ & & $--$\\

$\mathrm{Br}(B_{s}\to \mu e)$ & & $<1.4\times 10^{-8}$~\cite{bmue}\\

$\mathrm{Br}(B_{s}\to \tau e )$ & & $<2.8\times 10^{-5}$~\cite{pdg}\\

$\mathrm{Br}(B_{s}\to \mu \tau)$ & & $<2.2\times 10^{-5}$~\cite{pdg}\\

$\mathrm{Br}(B^+\to \tau^+ \nu)$ & & $(1.14\pm 0.27)\times
10^{-4}$~\cite{pdg} \\

$\mathrm{Br}(B\to D\tau \nu)/\mathrm{Br}(B\to Dl \nu)$ & &
($0.440 \pm 0.058 \pm 0.042)$~\cite{bdtaunu} \\

$\mathrm{Br}(B\to D^\star\tau \nu)/\mathrm{Br}(B\to D^\star l \nu)$ & &
($0.332 \pm 0.024 \pm 0.018)$~\cite{bdtaunu} \\

$\mathrm{Br}(B\to X_{s} \gamma)$ & & $(3.52\pm 0.25) \times
10^{-4}$~\cite{dmbd}\\

$\mathrm{Br}(t\to c h, u h)$ & & $ < 5.6\times 10^{-3}$~\cite{CMStuh}\\

\hline

\multicolumn{3}{|c|}{$\Delta F=2$}\\ \hline 

$|\epsilon_{K}|$ & & $(2.229 \pm 0.010)\times 10^{-3}$~\cite{pdg} \\

$\Delta M_{K}$ & & $(5.292 \pm 0.009)\times
10^{-3}~\mathrm{ps}^{-1}$~\cite{pdg}\\

$\Delta M_{D}$ & & $(2.37^{+0.66}_{-0.71}) \times
10^{-2}~\mathrm{ps}^{-1}$~\cite{pdg}\\
$\Delta M_{B_{d}}$ & & $(0.507 \pm
0.005)~\mathrm{ps}^{-1}$~\cite{dmbd}\\

$\Delta M_{B_{s}}$ & & $(17.77 \pm
0.12)~\mathrm{ps}^{-1}$~\cite{dmbs}\\ 

\hline

\end{tabular}
\end{center}
\label{tab:proc}
\caption{List of observables calculated by \code{} v2.5 and their
  measured values.}
\end{table}%

In summary, the basic features of \code{} v2.5 are:
\begin{itemize}
\item The program utilizes the most general $R$-parity conserving
  Lagrangian for the MSSM.  In addition to the standard soft breaking
  terms, it can accommodate for additional non-holomorphic trilinear
  soft-SUSY breaking terms,
\bea
A_l^{'IJ} H_i^{2\star} L_i^I R^J+
A_d^{'IJ} H_i^{2\star} Q_i^I D^J +
A_u^{'IJ} H_i^{1\star} Q_i^I U^J + \mathrm{H.c.} \;,
\eea
that do not appear in the minimal supergravity scenario but are
present in the most general softly broken supersymmetric effective
Lagrangian~\cite{Hall}.  These non-holomorphic terms can give rise to
sizable effects in Higgs-fermion couplings.
\item \code{} can read and accept without modifications most of the
  SLHA2-compatible output files produced by other public libraries
  calculating various aspects of the MSSM phenomenology.

\item There is no limit on the size of flavor violating parameters
  because the calculation does not rely on the MIA expansion.
  However, if the off-diagonal elements are larger than the diagonal
  ones, imaginary sfermion masses would be induced.  Complex ``mass
  insertions'' of the form
\bea
\delta^{IJ}_{QXY} &=& \frac{(M^2_{Q})^{IJ}_{XY}}{\sqrt{
    (M^2_{Q})^{II}_{XX} (M^2_{Q})^{JJ}_{YY} }}\;,
\label{eq:phys:massinsert}
\eea
($I,J$ denote quark flavors, $X,Y$ denote superfield chirality, and
$Q$ indicates either the up or down quark superfield sector, similarly
for slepton superfields) are taken as inputs, but they only serve to
conveniently parametrize the sfermion mass matrices.  \code{}
numerically calculates the exact tree-level spectrum and mixing
matrices, which are later used in loop calculations.

\item After calculating SUSY spectrum, \code{} performs the
  resummation of the chirally enhanced corrections (following the
  systematic approach of ref.~\cite{JRCRIV}), arising in the regime of
  large $\tan\beta$ and/or large trilinear soft sfermion mixing.  The
  values of the Yukawa couplings and the CKM matrix elements of the
  superpotential are calculated (by taking into account the threshold
  corrections) and are then used for the calculations of the SUSY loop
  contributions to flavor observables.  These chirally enhanced
  corrections also lead to flavor-changing neutral Higgs couplings and
  corrections to charged Higgs vertices which are implemented as well
  in the calculation of the amplitudes.

\item As an intermediate step parton-level form factors for quark and
  lepton 2-, 3- and 4-point Green functions are calculated.  They are
  later dressed in hadronic matrix elements (see Table~\ref{tab:green}
  in Sec.~\ref{sec:structure}) to obtain predictions for the physical
  quantities listed in Table~\ref{tab:proc}.  The set of Green's
  functions computed by \code{} as intermediate ``building blocks'' is
  quite universal and can be used for a calculation of various
  processes not yet implemented in \code{}.

\item The full list of the processes which can be calculated by
  \code{} v2.5 is given in Table~\ref{tab:proc}.
\end{itemize}

This article is organized as follows.  In Sec.~\ref{sec:mssmlag} we
discuss the conventions used for the MSSM parameters (a more explicit
description can be found in the manual of \code{} v1~\cite{SFLAV}).
Sec.~\ref{sec:structure} describes the internal structure of \code{},
the most important steps of the calculations and the file structure of
the library.  In Sec.~\ref{sec:init} we define the input parameters
and present the initialization sequence for \code.  Sec.~\ref{sec:ren}
discusses how the resummation of the chirally enhanced corrections to
all orders of perturbation theory is performed.  Routines for
calculating the flavor and CP observables collected in
Table~\ref{tab:proc} are described in Sec.~\ref{sec:proc}.  In
Sec.~\ref{sec:output} the output format for the quantities calculated
by \code{} is presented.  We conclude in Sec.~\ref{sec:summary} with a
summary of the presentation.  Appendix~\ref{app:inst} contains brief
instructions on how to install and run the \code{} package.
In appendices~\ref{app:code} and~\ref{app:infile} we provide templates
for initializing \code{} from within the program and using an external
file in the SLHA2 format~\cite{SLHA2}, respectively.  Both of these
templates produce the set of test results listed in
Appendix~\ref{app:output}.

\code{} can be downloaded from the following address\footnote{For an
  additional information, bug reports or any other questions related
  to the code please contact \code{} maintainer at the address {\tt
    janusz.rosiek@fuw.edu.pl}}:

\vspace*{0.3cm} \centerline{\webpage}

\section{Lagrangian and conventions}
\label{sec:mssmlag}

\code{} is capable of calculating physical observables within the most
general $R$-parity conserving MSSM, with one exception: currently it
assumes massless neutrinos (and no right neutrino and right sneutrino
fields in the Lagrangian~\cite{DEHARO}), so the PMNS mixing matrix
does not appear in any lepton and slepton couplings. Neutrino flavor
mixing and the PMNS matrix should be taken into account once new
experiments are able to identify the flavor of the neutrinos produced
in rare decays, but at present this is not experimentally feasible.
Still, over 100 Lagrangian parameters are taken as input to \code{}
and can be initialized independently.

\code{} has been in development since 1996, long before the Les
Houches Accord~\cite{SLHA} (SLHA), followed in 2008 by
SLHA2~\cite{SLHA2}, for common MSSM conventions was established.  By
the time SLHA2 became a commonly accepted standard, it was no longer
feasible to change the internal \code{} structure.  Thus, its internal
routines follow the conventions for the MSSM Lagrangian and Feynman
rules given in the earlier paper~\cite{PRD41}.  However, by default
\code{} can be initialized with a SLHA2 compatible set of parameters,
necessary translations are done in a way invisible for an user.

Actually, the choice of convention for the input parameters of \code{}
is a user-defined option.  If required, parameters can be also
initialized directly following the~\cite{PRD41} conventions.  The
choice between SLHA2 and ref.~\cite{PRD41} can be made by setting the
relevant control variable, as described in Sec.~\ref{sec:slhainp}.  In
Table~\ref{tab:slha} we summarize the (rather minor) differences
between the conventions of the extended SLHA2~\cite{SLHA2} and those
of ~\cite{PRD41}.

\begin{table}[htbp]
\begin{center}
\begin{tabular}{|c|c|}
\hline
SLHA2~\cite{SLHA2} & Ref.~\cite{PRD41}\\
\hline 
& \\[-4mm]
$\hat T_U$, $\hat T_D$, $\hat T_E$ & $-A_u^T$, $+A_d^T$, $+A_l^T$\\
$\hat m_{\tilde Q}^2$, $\hat m_{\tilde L}^2$ & $m_Q^2$, $m_L^2$ \\
$\hat m_{\tilde u}^2$, $\hat m_{\tilde d}^2$, $\hat m_{\tilde l}^2$ &
$(m_U^2)^T$, $(m_D^2)^T$, $(m_E^2)^T$ \\
${\cal M}_{\tilde u}^2$, ${\cal M}_{\tilde d}^2$ & $({\cal M}_U^2)^T$,
$({\cal M}_D^2)^T$ \\[1mm]
\hline
\end{tabular}
\end{center}
\caption{Comparison of~SLHA2~\cite{SLHA2} and Ref.~\cite{PRD41}
  conventions.}
\label{tab:slha}
\end{table}

One should note that in \code{} one can also use non-standard
trilinear scalar couplings, involving the complex conjugated Higgs
fields (sometimes called ``non-analytic'' or ``non-holomorphic''
$A$-terms).  In the notation of~\cite{PRD41} they read as:
\bea
A_l^{'IJ} H_i^{2\star} L_i^I E^J + A_d^{'IJ} H_i^{2\star} Q_i^I D^J +
A_u^{'IJ} H_i^{1\star} Q_i^I U^J + \mathrm{h.c.}
\label{eq:nhol}
\eea
Usually these couplings are not considered as they are not generated
in standard SUSY breaking models.  However, they are included
in~\code{} and by default initialized to zero.  Users may decide to
set them to some non-trivial values in order to check their impact on
rare decays phenomenology (loop corrections non-holomorphic $A$-terms
may lead to large flavor-changing neutral Higgs couplings).

In general, the parameter $\mu$, the soft-SUSY breaking Higgs-mass
term $m_{12}^2$, the gaugino mass parameters $M_{1,2,3}$, the soft
sfermion mass matrices and the trilinear soft couplings may be
complex.  Global rephasing of all fermion fields of the theory and of
one of the Higgs multiplets can render two of these parameters
real~\cite{POROSA}.  We choose them to be the gluino mass $M_3$ and
the Higgs mass term $m_{12}^2$.  The latter choice keeps the Higgs
vacuum expectation values (VEV) and, therefore, the parameter
$\tan\beta$ real at tree level.

\section{Structure of the code}
\label{sec:structure}

Calculations in \code{} take the following steps:

\noindent {\bf 1.  Parameter initialization.}  This is described in
details in Sec.~\ref{sec:init}.  Users can adjust the basic Standard
Model parameters according to latest experimental data and initialize
all (or the chosen subset of) supersymmetric soft masses and couplings
and Higgs sector parameters. The supersymmetric input parameters for
the \code{} must be given at the SUSY scale and program offers no
internal routines for evolving them to other scales.  At this step
also various QCD- and hadronic-related quantities, like e.g. hadronic
matrix element values, can be adjusted.

\noindent {\bf 2.  Calculation of the physical masses and the mixing
  angles.}  After setting the input parameters, \code{} calculates the
eigenvalues of the mass matrices of all MSSM particles and their
mixing matrices at tree level.  Diagonalization is done numerically
without any approximations.

\noindent {\bf 3.  Resummation of the chirally enhanced effects.} In
the regime of large $\tan\beta$ and/or large trilinear SUSY breaking
terms, large chirally enhanced corrections to Yukawa couplings and CKM
matrix elements arise.  \code{} v2 can perform resummation of these
corrections to all orders of perturbation theory.  After calculating
threshold corrections, the Yukawa couplings and CKM elements of the
superpotential (i.e. the ``bare'' parameters) are determined.  Using
these quantities the chirally enhanced effects are calculated and
absorbed into effective Higgs-fermion and
fermion-sfermion-gaugino(higgsino) vertices.  Using these vertices in
the calculation of flavor observables, all chirally enhanced
corrections are automatically taken into account.  The level of
resummation (no resummation, approximate analytical resummation in the
decoupling limit, iterative numerical resummation) is a user defined
option.

\noindent {\bf 4.  Calculation of the Wilson coefficients at the SUSY
  scale}.  The one-loop Wilson coefficients of the effective operators
required for a given process are calculated using the sfermion mixing
matrices and the physical masses as input.  Again, the formulae used
in the code do not rely on any approximations, such as the MIA
expansion.  In the current version, \code{} calculates Wilson
coefficients generated by the diagrams listed in
Table~\ref{tab:green}.  All Wilson coefficients are calculated at the
energy scale assumed to be the average mass of SUSY particles
contributing to a given process or the top quark scale.

\begin{table}[htbp]
\label{tab:green}
\begin{center}
\begin{tabular}{|cp{1mm}|p{1mm}cp{1mm}|p{1mm}c|}
\hline
Box &&& Penguin &&& Self energy \\[2mm] \hline\hline
$dddd$ &&& $Z\bar d d$, $\gamma \bar d d$, $g \bar d d$ &&& $d$-quark \\
$uuuu$ &&& $H_i^0 \bar d d$, $A_i^0 \bar d d$ &&& $u$-quark \\
$ddll$ &&& $H_i^0 \bar u u$, $A_i^0 \bar u u$ &&& charged lepton $l$\\
$dd\nu\nu$ &&& $\gamma\bar l l$ &&&  \\
\hline
\end{tabular}\\[2mm]
\caption{One loop parton level diagrams implemented in \code{}.}
\end{center}
\end{table}

It is important to stress that routines of \code{} calculating form
factors accept fermion generation indices as input parameters.  Thus
in Table~\ref{tab:green} $d$ and $u$, $l$ and $\nu$ denote quarks or
leptons of {\em any} generation.  Hence, the actual number of
amplitudes which can be calculated using combinations of these form
factors is much larger than used by the rare decay rates currently
implemented fully in \code, opening possibility for further
developments of the library.

\noindent {\bf 5.  Strong corrections.}  In the final step \code{}
performs (when necessary) the QCD evolution of the Wilson coefficients
from the high scale (SUSY or top quark mass scale) to the low energy
scale appropriate for a given decay, calculates the relevant hadronic
matrix elements, and returns predictions for physical quantities.  The
formulae for QCD and hadronic corrections are primarily based on
calculations performed in the SM and supplemented, when necessary,
with contributions from non-standard operators which usually are
neglected in the SM, because they are suppressed by powers of the
light quark Yukawa couplings.  This part of \code{} is based on
analyses published by other authors, whereas points 1-4 are
implemented using our own calculations.  The accuracy of strong
corrections differ from process to process, from negligible or small
(leptonic EDM, ``gold-plated'' decay modes $K\ra \pi\bar \nu
\nu$~\cite{BB98}) to order of magnitude uncertainties (unknown long
distance contributions to $\Delta m_K$ or $\Delta m_D$).  Even in the
case of large QCD uncertainties, the result of the calculation
performed by \code{} can be of some use.  Flavor violation in the
sfermion sector can lead to huge modifications of many observables,
sometimes by several orders of magnitude, so that comparison with
experimental data can help to constrain the soft flavor-violating
terms even if the strong corrections are not very well known.

In Table~\ref{tab:files} we list the files included in \code{} library
with a brief description of their content and purpose.  Most of the
2-, 3- and 4-point Green functions are calculated for vanishing
external momenta (exception are up-quark self energies and Higgs-up
quark 3-point functions where Higgs boson and top quark masses are not
small enough to be neglected).  As mentioned before, by ``$u$ quark''
and ``$d$ quark'' we mean all generations of quarks.  In addition to
files listed in Table~\ref{tab:files}, the library contains the master
driver files {\tt susy\_flavor\_file.f} and {\tt susy\_flavor\_prog.f}
which illustrate the proper initialization sequence for \code{}
parameters and produce a sample of results for the implemented
observables.

\begin{table}[htbp]
\noindent
{\small
\begin{tabular}{rp{0mm}p{13cm}} 
{\tt b\_fun.f:} && general 2-point loop functions \\

{\tt bsg\_nl.f:} && formulae for $\mathrm{Br}(B\ra X_s \gamma)$,
including QCD corrections \\

{\tt cdm\_q.f:} && $u$- and $d$-quark chromoelectric dipole moments \\

{\tt cdm\_g.f:} && gluon chromoelectric dipole moment \\

{\tt c\_fun.f:} && general 3-point loop functions \\

{\tt c\_fun\_exp.f:} && 3-point functions $c_0, c_{11}, c_{12}$
expanded in external momenta \\

{\tt cd\_fun.f:} && 3-, 4- and some 5-point loop functions at
vanishing external momenta \\

{\tt db\_fun.f:} && derivatives of general 2-point loop functions \\

{\tt dd\_gamma.f:} && $d$ quark-$d$ quark-photon 1-loop triangle
diagram \\

{\tt ddg\_fun.f:} && general gauge boson-fermion-fermion 1-loop
triangle diagram \\

{\tt dd\_gluon.f:} && $d$ quark-$d$ quark-gluon 1-loop triangle
diagram \\

{\tt dd\_ll.f:} && $d$ quark-$d$ quark-lepton-lepton 1-loop box
diagram \\

{\tt dd\_mix.f:} && 4-$d$ quark 1-loop box diagram \\

{\tt dd\_vv.f:} && $d$ quark-$d$ quark-neutrino-neutrino 1-loop box
diagram \\

{\tt d\_self0.f:} && full $d$-quark self-energy \\

{\tt edm\_q.f:} && $u$- and $d$-quark electric dipole moments \\

{\tt eisch1.f:} && auxiliary numerical routine - hermitian matrix
diagonalization \\

{\tt l\_self0\_dlim.f:} && routines for the various decompositions of
the lepton self energies \\

{\tt ll\_gamma.f:} && lepton-lepton-photon 1-loop triangle diagram \\

{\tt mh\_diag.f:} && diagonalization of tree level mass matrices,
approximate 2-loop Higgs mass~$m_h$ \\

{\tt mh\_init.f:} && initialization of MSSM parameters \\

{\tt phen\_2l.f:} && formulae for $Br(\mu\to e\gamma)$,
$\mathrm{Br}(\tau\ra \mu\gamma, e\gamma)$, lepton $g-2$ anomaly and
EDMs \\

{\tt phen\_2q.f:} && formulae for $Br(K_L^0\ra \pi^0 \bar\nu \nu)$,
$\mathrm{Br}(K^+\ra \pi^+ \bar\nu \nu)$, $\mathrm{Br}(B_{s(d)}\ra
l^+l^-)$, $\mathrm{Br}(B\ra \tau\nu, D\tau\nu)$, $\mathrm{Br}(t\ra
uh,ch)$, $\mathrm{Br}(t\ra ug,cg)$ and neutron EDM \\

{\tt phen\_4q.f:} && formulae for the meson mixing observables:
$\Delta m_K$, $\epsilon_K$, $\Delta m_D$, $\Delta m_{B_{d(s)}}$\\

{\tt qcd\_fun.f:} && auxiliary QCD calculations - running $\alpha_s$,
running quark masses etc.  \\

{\tt q\_self0\_dlim.f:} && routines for the various decompositions of
the $u$- and $d$-quark self energies\\

{\tt rombint.f:} && auxiliary numerical routine - Romberg numerical
integration \\

{\tt sff\_fun.f:} && general scalar-fermion-fermion 1-loop triangle
diagram \\

{\tt sflav\_io.f:} && input/output routines for the SLHA2 data format
\\

{\tt sflav\_main.f:} && main routine calculating all physical
observables \\

{\tt suu\_vert.f:} && CP-even neutral Higgs boson-$u$ quark-$u$ quark
1-loop triangle diagram\\

{\tt u\_self0.f:} && $u$-quark self-energy \\

{\tt uu\_gluon.f:} && $u$ quark-$u$ quark-gluon 1-loop triangle diagram
\\

{\tt uu\_mix.f:} && 4-$u$ quark 1-loop box diagram \\

{\tt vegas.f:} && auxiliary numerical routine - Vegas Monte Carlo
integration \\

{\tt vf\_def.f:} && definitions of fermion tree-level vertices \\

{\tt vg\_def.f:} && definitions of gauge boson tree-level vertices \\

{\tt vh\_def.f:} && definitions of Higgs boson tree-level vertices \\

{\tt yuk\_ren.f:} && chiral corrections to the Yukawa couplings and
CKM matrix \\

{\tt zdd\_vert0.f:} && $Z$ boson-$d$ quark-$d$ quark 1-loop triangle
diagram

\end{tabular}
}
\vspace*{-4mm}
\caption{List of files included in \code{} library.
  \label{tab:files}}
\end{table}

\section{Parameter initialization in \code}
\label{sec:init}

Apart from initialization routines used by \code{} and their arguments
we list here the FORTRAN common blocks storing the most important
program data (other common blocks serve for the internal purposes and
usually do not need to be accessed by users).  As mentioned in the
previous section, supersymmetric input parameters should be given at
the SUSY scale (only for some SM parameters, like running quark
masses, the input scale is user defined).

By default, \code{} uses the following implicit type declaration in
all routines:\\[2mm]
{\tt implicit double precision (a-h,o-z)}\\[2mm]
so that all variables with names starting from {\tt a} to {\tt h} and
from {\tt o} to {\tt z} are automatically defined as {\tt double
  precision} and those with names starting from {\tt i} to {\tt n} are
of {\tt integer} type.  In what follows we indicate variables that do
not obey this rule.  Such variables are always listed in explicit type
statements inside the procedures.  Complex parameters are declared in
\code{} as {\tt double complex} type.  Mass parameters are always
given in~GeV.

\code{} provides two ways of initializing the input parameters.
Firstly, they can be read from the file {\tt susy\_flavor.in}.  The
structure of this file follows the SLHA2 convention~\cite{SLHA2}, with
optional extensions which we describe in Sec.~\ref{sec:slhainp}.
Initializing parameters in the input file does not require a detailed
knowledge of the program internal structure.  This option, as it
requires a disk file access for each parameter set may not be most
efficient for scans over the MSSM parameter space.  Therefore, \code{}
provides also a set of routines designed to initialize parameters
defined in the program, which can be used to prepare programs that
scan over large parameter sets.  As described in
Sec.~\ref{sec:proginp}, these routines require more care, as they
should be initialized in proper order, i.e. first the gauge sector,
then the fermion sector, Higgs sector, and at the end SUSY sectors
(the initialization sequences for the gaugino, slepton and squark
sectors are independent).

An examples of an initialization sequence for \code, illustrating both
options mentioned above, is presented Appendix~\ref{app:code}.  The
sample input file {\tt susy\_flavor.in} is given in
Appendix~\ref{app:infile}.  Test output generated for parameters used
in Appendices~\ref{app:code} and~\ref{app:infile} is enclosed in
Appendix~\ref{app:output}.

\subsection{Variables  controlling particle content}
\label{sec:debug}

\code{} v2.5 allows to separately switch contributions from various
MSSM sectors on or off.  Such a feature is useful to understand the
relative size of their effects for each of the calculated
processes. The relevant control variables can be set by the following
FORTRAN statement at the beginning of the driver program:

 {\tt call set\_active\_sector(ih,ic,in,ig)},

\noindent where the variables {\tt ih, ic, in} and {\tt ig} can take
values $0$ or $1$ and they control, respectively, the inclusion in the
total result the diagrams with gauge and Higgs bosons, charginos,
neutralinos and gluinos exchanged in the loops. Note that diagrams
with Higgs and gauge bosons circulating in loops are always added
together and currently cannot be disentangled, so setting {\tt ih=1,
  ic=in=ig=0} does not reproduce the SM result. Also for $\Delta F=2$
processes, where mixed box diagrams with both neutralino and gluino in
the loop exist, such diagrams are included only if both {\tt in=ig=1}.

Obviously, by default, if no call to {\tt set\_active\_sector} is
made, all control variables are assumed to be equal 1, so that all
contributions are included.

\subsection{Parameter initialization from the input file}
\label{sec:slhainp}

The input parameters for \code{} can be set by the editing appropriate
entries of the file {\tt susy\_flavor.in} and subsequently calling the
subroutine {\tt sflav\_input}, which reads the input file, stores 
the MSSM Lagrangian parameters in FORTRAN common blocks and calculates
tree-level physical masses and mixing matrices.  After calling {\tt
  sflav\_input}, all physical observable described in
Sec.~\ref{sec:proc} can be calculated.  The input file {\tt
  susy\_flavor.in} is written in the SLHA2 format, with some
extensions which we list below.

The initialization proceeds as follows. Before reading the input file,
all parameters are set to some initial values. In version 2.50 they
are:
\begin{itemize}
\item basic SM parameters\\
\begin{tabular}{lp{5mm}lp{5mm}l}
$\alpha_{em}(M_Z) = 1/127.934$ && $M_Z = 91.1876$ GeV &&
  $s_W^2(\rm{MSBar}) = 0.23116$ \\
$\alpha_s(M_Z) = 0.1172$ && $M_W = 80.398$ GeV \\
\end{tabular}
\item quark-related parameters\\[2mm]
\begin{tabular}{lp{5mm}lp{5mm}l}
running quark masses && pole fermion masses && CKM parameters \\[2mm]
$m_u(2~\rm{GeV}) = 2.15$ MeV && $m_t = 173.5$ GeV && $\lambda=0.2258$\\
$m_d(2~\rm{GeV}) = 4.7$ MeV && $m_e = 0.5109989$ MeV &&  $A=0.808$ \\
$m_s(2~\rm{GeV}) = 93.5$ MeV && $m_\mu = 105.658$ MeV && $\bar\rho=
0.177$ \\
$ m_c(m_c) = 1.275$ GeV && $m_\tau = 1.77684$ GeV && $\bar\eta= 0.36$
\\
$m_b(m_b) = 4.18$ GeV &&
\end{tabular}
\item all MSSM mass parameters ($\mu$, gaugino and sfermion masses,
  trilinear $A$ terms) are set to 0.  $\tan\beta$ and the CP-odd Higgs
  mass $M_A$, which we use as the input parameters for the Higgs
  sector, are also set to 0.
\item hadronic-related parameters (QCD scales and effective
  coefficients, hadronic matrix elements etc.) are set to values
  described in Sections~\ref{sec:gminus2}--\ref{sec:bbmix}. Their
  compact list is given in {\tt Block SFLAV\_HADRON} in
  Appendix~\ref{app:infile}.
\end{itemize}

Subsequently, the input Blocks are read from the file {\tt
  susy\_flavor.in} in the following order: {\tt SOFTINP}, {\tt
  SMINPUTS}, {\tt VCKMIN}, {\tt MINPAR} ($\tan\beta$ only, other
entries ignored), {\tt EXTPAR}, {\tt IMEXTPAR}, {\tt MSL2IN}, {\tt
  IMMSL2IN}, {\tt MSE2IN}, {\tt IMMSE2IN}, {\tt TEIN}, {\tt IMTEIN},
{\tt TEINH}, {\tt IMTEINH}, {\tt MSQ2IN}, {\tt IMMSQ2IN}, {\tt
  MSU2IN}, {\tt IMMSU2IN}, {\tt MSD2IN}, {\tt IMMSD2IN}, {\tt TUIN},
{\tt IMTUIN}, {\tt TUINH}, {\tt IMTUINH}, {\tt TDIN}, {\tt IMTDIN},
{\tt TDINH}, {\tt IMTDINH}, {\tt SFLAV\_HADRON}.

In principle the presence of {\em any} Block is optional - if some
Block is absent, the program falls back to default parameter values
listed above.  Obviously, at least flavor-diagonal SUSY mass
parameters have to be defined, otherwise the vanishing default masses
will cause the crash of the program. If a parameter is multiply
defined in several Blocks (for example left slepton mass parameters in
{\tt Block EXTPAR} and later in {\tt Blocks MSL2IN, IMMSL2IN}), the
value from Block read as latest in the list above overwrites (without
warning!)  the values from preceding Blocks. Blocks do not need to be
complete, i.e. to contain all entries described in SLHA2 specification
- it is sufficient to define minimal set of parameters relevant for
given problem, others would be filled with default values.

Comparing to standard SLHA2 conventions, \code{} uses following
extensions:

\noindent 1. We define an optional {\tt Block SOFTINP} defining choice
of input conventions.  If such block is not present, program assumes
default values of control variables: \\[2mm]
\begin{tabular}{lcp{125mm}}
  Variable value && Sfermion sector parametrization \\[2mm]
  ${\tt iconv}=1$ && {\bf default:} MSSM parameters defined in SLHA2
  conventions. \\
  ${\tt iconv}=2$ && MSSM parameters defined in conventions of
  Ref.~\cite{PRD41}.\\[2mm]
  ${\tt input\_type}=1$ && off-diagonal soft terms are given as
  dimensionless mass insertions.\\
  ${\tt input\_type}=2$ && {\bf default:} sfermion soft terms given as
  absolute dimensionful values. \\[2mm]
  ${\tt ilev}=0$ && no resummation of chirally enhanced corrections,
  all SUSY contributions are strictly taken at the 1-loop level. \\
  ${\tt ilev}=1$ && resummation of chirally enhanced corrections 
  performed with the use of analytical formulae valid in the decoupling 
  limit $M_{SUSY}\gg v_1,v_2$. \\
  ${\tt ilev}=2$ && {\bf default:} resummation of chirally enhanced
  corrections performed using the numerical iterative solutions for
  bare Yukawa couplings and CKM matrix elements. \\
\end{tabular}

\noindent 2. \code{} uses two non-standard (comparing to SLHA2)
entries of {\tt Block SMINPUTS}. Entry 30 is used to define $M_W$ and
entry 31 to define $s_W^2$ in MSbar renormalization scheme.

\noindent 3.  Following the SLHA2 convention, full sfermion soft mass
matrices can be defined in the {\tt MSL2IN}, {\tt MSE2IN}, {\tt
  MSQ2IN}, {\tt MSD2IN}, {\tt MSU2IN} and {\tt IMMSL2IN}, {\tt
  IMMSE2IN}, {\tt IMMSQ2IN}, {\tt IMMSD2IN}, {\tt IMMSU2IN} blocks.
The {\tt input\_type} parameter in the {\tt SOFTINP} block defines the
dimension of the off-diagonal terms.  If ${\tt input\_type}=1$, the
off-diagonal entries given in {\tt susy\_flavor.in} are assumed to be
dimensionless mass insertions $\delta^{IJ}_X$ and the actual flavor
violating sfermion soft mass terms are calculated as
\bea
(m_X^2)_{IJ} &=& (m_X^2)_{JI}^\star = \delta^{IJ}_{X}
\sqrt{(m_X^2)_{II} (m_X^2)_{JJ}}\; ,
\label{eq:midef}
\eea
where $X=L,E,Q,U,D$ and $I,J$ enumerate superpartners of the
mass-eigenstates quarks.

\noindent 4. The blocks {\tt TEIN, TDIN, TUIN} and {\tt IMTEIN,
  IMTDIN, IMTUIN} define the full trilinear SUSY breaking terms. They
are in general not hermitian and one is required to define all
entries.  Again the parameter {\tt input\_type} defines the format and
dimension of the off-diagonal terms.  If ${\tt input\_type}=1$, then
all relevant {\tt susy\_flavor.in} entries are treated as
dimensionless numbers and expanded to the full trilinear SUSY breaking
terms as:
\bea
A_l^{IJ} &=& \delta^{IJ}_{LLR} \, \left((m_L^2)_{II}
(m_E^2)_{JJ}\right)^{\frac{1}{4}}\; ,\nonumber\\
A_d^{IJ} &=& \delta^{IJ}_{DLR} \, \left((m_Q^2)_{II}
(m_D^2)_{JJ}\right)^{\frac{1}{4}}\; ,\nonumber\\
A_u^{IJ} &=& \delta^{IJ}_{ULR} \, \left((m_Q^2)_{II}
(m_U^2)_{JJ}\right)^{\frac{1}{4}}\; .
\label{eq:lrmidef}
\eea
Note that the $A$-terms are normalized to the diagonal sfermion
masses, not to the diagonal trilinear terms, and that in
eq.~(\ref{eq:lrmidef}) for simplicity we use $(m_Q^2)_{II}$ as the
diagonal mass scale for both up and down left squark fields (related
by the CKM rotation).

\noindent 5.  The ``non-holomorphic'' LR mixing terms of
eq.~(\ref{eq:nhol}) are not included in the SLHA2 specification of the
MSSM parameters.  They can be defined if necessary in blocks {\tt
  TEINH, TDINH, TUINH} and {\tt IMTEINH, IMTDINH, IMTUINH}. If such
blocks are not present, all such terms are set to 0.  As standard LR
mixing terms, non-holomorphic ones are also not hermitian in general.
Again depending on the value of {\tt input\_type} they can be given as
dimensionful or dimensionless.  In the second case (${\tt
  input\_type}=1$) the dimensionful non-holomorphic terms are
calculated in a way analogous to eq.~(\ref{eq:lrmidef}).

\subsection{Parameter initialization inside the program}
\label{sec:proginp}

\code{} input parameters can be initialized directly inside the driver
program using the set of routines described below.  Before the proper
initialization sequence, the user can set the {\tt iconv}
variable value to choose the input convention:\\[2mm]
\begin{tabular}{lp{1mm}l}
{\tt common/sf\_cont/eps,indx(3,3),iconv} \\
\hskip 11mm {\tt iconv=1} && SLHA2~\cite{SLHA2} input conventions \\
\hskip 11mm {\tt iconv=2} && \cite{PRD41} input conventions\\
\end{tabular}\\[2mm]
After choosing the input conventions, one should subsequently
initialize the gauge, matter fermion, Higgs, SUSY fermion and sfermion
sectors (exactly in this order), using the procedures described in
detail in the following sections.

\subsubsection{Gauge sector}

As input, \code{} takes the gauge boson masses ($M_W, M_Z$) and the
gauge coupling constants (electromagnetic and strong) at the $M_Z$
scale.   They are initialized by: \\[2mm]
{\small
\begin{tabular}{lp{95mm}}
 Routine and arguments & Purpose and MSSM parameters \\[2mm]
{\tt vpar\_update(zm,wm,alpha\_{em},st2)} & Sets electromagnetic sector
parameters \\
\hskip 11mm {\tt zm} & $M_Z$, $Z$ boson mass\\
\hskip 11mm {\tt wm} & $M_W$, $W$ boson mass\\
\hskip 11mm {\tt alpha\_{em}} & $\alpha_{em}(M_Z)$, QED coupling at
$M_Z$ scale\\
\hskip 11mm {\tt st2} & $s_W^2$ in MSBar scheme\\[2mm]
{\tt lam\_fit(alpha\_s)} & Sets $\alpha_s(M_Z)$ and $\Lambda_{QCD}$
for 4-6 flavors at the NNLO level \\
{\tt lam\_fit\_nlo(alpha\_s)} & Sets $\alpha_s(M_Z)$ and
$\Lambda_{QCD}$ for 4-6 flavors at the NLO level \\
\hskip 11mm {\tt alpha\_s} & $\alpha_s(M_Z)$, strong coupling at
$M_Z$ scale\\
\end{tabular}\\[2mm]
}

\subsubsection{Matter fermion sector}
\label{sec:fmass}

\code{} assumes that neutrinos are massless.  The pole masses of the
charged leptons are initialized in the file\, {\tt sflav\_io.f} in the
routine {\tt sflav\_defaults} and can be adjusted changing the values
given there.
In the quark sector the most important input parameters are the
running top and bottom masses at a given renormalization scale and the
CKM angles and phase.  They can be set by:\\[2mm] {\small
\begin{tabular}{lp{77mm}}
Routine and arguments & Purpose and MSSM parameters \\[2mm]
{\tt init\_fermion\_sector(alpha\_s,tm,tsc,bm,bsc)} & Sets running top
and bottom quark mass \\
\hskip 11mm {\tt alpha\_s} & $\alpha_s(M_Z)$, strong coupling at
$M_Z$ scale\\
\hskip 11mm {\tt tm,tsc} & $m_t(\mu_t)$, running
$\overline{\mathrm{MS}}$ top quark mass\\
\hskip 11mm {\tt bm,bsc} & $m_b(\mu_b)$, running
$\overline{\mathrm{MS}}$ bottom quark mass\\[2mm]
{\tt ckm\_init(s12,s23,s13,delta) } & Option 1: initialization of the
CKM matrix \\
\hskip 11mm {\tt s12,s23,s13} & $\sin\theta_{12}, \sin\theta_{23},
\sin\theta_{13}$, sines of the CKM angles \\
\hskip 11mm {\tt delta} & $\delta$, the CKM phase in radians\\
{\tt ckm\_wolf(alam,a,rhobar,etabar) } & Option 2: initialization of
the CKM matrix \\
\hskip 11mm {\tt alam,a,rhobar,etabar} & Wolfenstein parameters
$\lambda,A.\bar\rho,\bar\eta$ \\
\end{tabular}\\[2mm]
} 
The light quark masses can be also adjusted by changing values which
are set in the routine {\tt sflav\_defaults}.

\subsubsection{Higgs sector}

Following the common convention, we take the Higgs mixing parameter
$\mu$, the CP-odd Higgs boson mass $M_A$, and the ratio of vacuum
expectation values $\tan\beta=v_2/v_1$ as the input parameters (in
order to calculate values of Higgs mass terms in the Lagrangian, one
needs to set also the $\mu$ parameter already here): \\[2mm]
\noindent {\small \begin{tabular}{lp{1mm}p{132mm}}
    \multicolumn{3}{l}{\tt subroutine
      init\_higgs\_sector(pm,tb,amu,ierr)}\\[2mm]
    Argument && MSSM parameters \\[2mm]

    {\tt pm} && CP-odd Higgs mass $M_A$\\

    {\tt tb} && Ratio of Higgs VEVs, $\tan\beta=\frac{v_2}{v_1}$\\

    {\tt amu} && Higgs mixing parameter $\mu$ (complex)\\

    {\tt ierr} && output error code: $ierr\neq 0$ if Higgs sector 
initialization failed \\[2mm]
\end{tabular}
}

\subsubsection{Sfermion sector}  
\label{sec:sferinit}

\code{} uses two subroutines to initialize sfermion parameters, {\tt
  init\_slepton\_sector} and {\tt init\_squark\_sector}.  They accept
as input diagonal masses and off-diagonal dimensionless mass
insertions, expanded later to entries of the soft mass matrices as
defined by eqs.~(\ref{eq:midef}), (\ref{eq:lrmidef}) (this is only a
choice of parametrization and does not lead to any loss of
generality).  The sfermion initialization routines have the following
arguments: \\[2mm]
\noindent {\small \begin{tabular}{lp{1mm}p{132mm}}
    \multicolumn{3}{l}{\tt subroutine
      init\_squark\_sector(sql,squ,sqd,asu,asd,sqmi\_l,sumi\_r,sdmi\_r, }\\
    &&{\tt sumi\_lr,sdmi\_lr,sumi\_lrp,sdmi\_lrp,ierr) }\\[2mm]

  Argument && MSSM parameters \\[2mm]

  {\tt sql} && Array of the diagonal left-handed down-squark masses
  $(m_D^2)_{LL}^{II}={\tt sql(I)}^2$, $I=1\ldots 3$\\

  {\tt squ} && Array of the diagonal right-handed up-squark masses
  $(m_U^2)_{RR}^{II}={\tt squ(I)}^2$, $I=1\ldots 3$\\

  {\tt sqd} && Array of the diagonal right-handed down-squark masses
  $(m_D^2)_{RR}^{II}={\tt sqd(I)}^2$, $I=1\ldots 3$\\

  {\tt sqmi\_l} && Array of the off-diagonal left-handed down squark mass
  insertions $(\delta_D)_{LL}^{12} = {\tt sqmi\_l(1)}$,
  ${\delta_D}_{LL}^{23} = {\tt sqmi\_l(2)}$, $(\delta_D)_{LL}^{13} =
  {\tt sqmi\_l(3)}$ (complex parameters); remaining down LL mass
  insertions are initialized via hermitian conjugation;  up LL mass
  matrix obtained via $SU(2)$ relation\\

  {\tt sumi\_r} && Array of the off-diagonal right-handed up-squark mass
  insertions $(\delta_U)_{RR}^{12} = {\tt sumi\_r(1)}$,
  $(\delta_U)^{23}_{RR} = {\tt sumi\_r(2)}$, $(\delta_U)^{13}_{RR} =
  {\tt sumi\_r(3)}$ (complex parameters); remaining up RR mass
  insertions  are initialized via hermitian conjugation\\

  {\tt sdmi\_r} && Array of the off-diagonal right-handed down-squark mass
  insertions $(\delta_D)^{12}_{RR} = {\tt sdmi\_r(1)}$,
  $(\delta_D)^{23}_{RR} = {\tt sdmi\_r(2)}$, $(\delta_D)^{13}_{RR} =
  {\tt sdmi\_r(3)}$ (complex parameters); remaining down RR mass
  insertions are initialized via hermitian conjugation\\

  {\tt sumi\_lr} && Matrix with the standard (holomorphic) up-squark
  trilinear LR mass insertions $(\delta_U)^{IJ}_{LR} = {\tt
    sumi\_lr(I,J)}$, $I,J=1\ldots 3$ (complex parameters)\\

  {\tt sdmi\_lr} && Matrix with the standard (holomorphic) down-squark
  trilinear LR mass insertions $(\delta_D)^{IJ}_{LR} = {\tt
    sdmi\_lr(I,J)}$, $I,J=1\ldots 3$ (complex parameters)\\

  {\tt sumi\_lrp} && Matrix with the non-holomorphic up-squark
  trilinear LR mass insertions $(\delta_U')^{IJ}_{LR} = {\tt
    sumi\_lrp(I,J)}$, $I,J=1\ldots 3$ (complex parameters)\\

  {\tt sdmi\_lrp} && Matrix with the non-holomorphic down-squark
  trilinear LR mass insertions $(\delta_D')^{IJ}_{LR} = {\tt
    sdmi\_lrp(I,J)}$, $I,J=1\ldots 3$ (complex parameters)\\

  {\tt ierr} && output error code: $ierr\neq 0$ if squark sector
  initialization failed (negative physical squark mass$^2$) \\

\end{tabular}\\[2mm]
}

\noindent {\small
  \begin{tabular}{llp{132mm}}
    \multicolumn{3}{l}{\tt subroutine
      init\_slepton\_sector(sll,slr,slmi\_l,slmi\_r,slmi\_lr,slmi\_lrp,ierr)}\\[2mm]

    Argument && MSSM parameters \\[2mm]

    {\tt sll} && Array of the diagonal left-handed slepton masses
    $(m_L^2)_{LL}^{II}={\tt sll(I)}^2$, $I=1\ldots 3$\\

    {\tt slr} && Array of the diagonal right-handed slepton masses
    $(m_L^2)_{RR}^{II}$=${\tt slr(I)}^2$, $I=1\ldots 3$\\

    {\tt slmi\_l} && Array of the off-diagonal left-handed slepton mass
    insertions $(\delta_L)^{12}_{LL} = {\tt slmi\_l(1)}$,
    $(\delta_L)^{23}_{LL} = {\tt slmi\_l(2)}$, $(\delta_L)^{13}_{LL} =
    {\tt slmi\_l(3)}$ (complex parameters); remaining LL mass insertions
    are initialized via hermitian conjugation\\

    {\tt slmi\_r} && Array of the off-diagonal right-handed slepton mass
    insertions $(\delta_L)^{12}_{RR} = {\tt slmi\_r(1)}$,
    $(\delta_L)^{23}_{RR} = {\tt slmi\_r(2)}$, $(\delta_L)^{13}_{RR} =
    {\tt slmi\_r(3)}$ (complex parameters); the remaining RR mass insertions
    are initialized via hermitian conjugation\\

    {\tt slmi\_lr} && Matrix with the standard (holomorphic) slepton
    trilinear LR mass insertions $(\delta_L)^{IJ}_{LR} = {\tt
      slmi\_lr(I,J)}$, $I,J=1\ldots 3$ (complex parameters)\\

    {\tt slmi\_lrp} && Matrix with the non-holomorphic slepton trilinear
    LR mass insertions $(\delta_L')^{IJ}_{LR} = {\tt slmi\_lrp(I,J)}$,
    $I,J=1\ldots 3$ (complex parameters)\\

    {\tt ierr} && output error code: $ierr\neq0$ if slepton sector
    initialization failed (negative physical slepton mass$^2$) \\

\end{tabular}
}

\subsubsection{Supersymmetric fermion sector}

Initialization is done by the routine {\tt init\_ino\_sector}:\\[2mm]
\noindent {\small \begin{tabular}{lp{1mm}p{132mm}}
    \multicolumn{3}{l}{\tt subroutine
      init\_ino\_sector(gm1,gm2,gm3,amu,tb,ierr)}\\[2mm]
    Argument && MSSM parameters \\[2mm]
    
    {\tt gm1,gm2} && $U(1), SU(2)$ gaugino masses (complex) \\
    
    {\tt gm3} && $SU(3)$ gaugino mass \\

    {\tt tb} && $\tan\beta=\frac{v_2}{v_1}$, the ratio of
    Higgs VEVs \\

    {\tt amu} && the Higgs mixing parameter $\mu$ (complex) \\

    {\tt ierr} && output warning code: $ierr\neq 0$ for
    chargino or  neutralino  lighter than $M_Z/2$ \\[2mm]

  \end{tabular}\\[2mm]
} If one sets $M_1=0$ in the call to {\tt init\_ino\_sector} then the
GUT-derived relation $M_1 = \frac{5}{3}\tan^2\theta_W M_2$ is used for
$M_1$.

\subsection{Tree-level physical masses and mixing angles}
\label{sec:commons}

After performing the full initialization sequence in \code{}, all the
MSSM Lagrangian parameters, physical tree-level particle masses (with
the exception of the running quark masses), and mixing matrices are
calculated and stored in common blocks.  If necessary, they can be
directly accessed and modified.  Note, however, that after any
modifications of the Lagrangian parameters, relevant procedures
calculating physical masses and mixing angles have to called again.
In Table~\ref{tab:mssmpar} we list the important blocks storing MSSM
parameters.  Common blocks containing masses and mixing angles are
listed in Table~\ref{tab:eigenmass}.

\begin{table}[htbp]
  {\small \begin{center} \begin{tabular}{lp{92mm}}
        Common block and variables & Lagrangian parameters \\[2mm]
  
        \multicolumn{2}{l}{\tt
          common/vpar/st,ct,st2,ct2,sct,sct2,e,e2,alpha,wm,wm2,zm,zm2,pi,sq2}
        \\

        \hskip 9mm {\tt st,ct,st2,ct2,sct,sct2} & Weinberg angle functions,
        respectively $s_W$, $c_W$, $s_W^2$, $c_W^2$, $s_Wc_W$, $s_W^2 c_W^2$\\

        \hskip 9mm {\tt e,e2,alpha} & electric charge powers at $M_Z$ scale:
        $e$, $e^2$, $\alpha_{em}$ \\

        \hskip 9mm {\tt wm,wm2,zm,zm2} & gauge boson masses: $M_W$, $M_W^2$,
        $M_Z$, $M_Z^2$\\

        \hskip 9mm {\tt pi,sq2} & numerical constants, $\pi$ and $\sqrt{2}$
        \\[2mm]

        \multicolumn{2}{l}{\tt common/hpar/hm1,hm2,hm12,hmu} \\

        \hskip 9mm {\tt hm1,hm2} & soft Higgs masses $m_{H_1}^2,
        m_{H_2}^2$\\

        \hskip 9mm {\tt hm12} & soft Higgs mixing parameter $m_{12}^2$\\

        \hskip 9mm {\tt hmu} & Higgs mixing parameter $\mu$ (complex)\\[2mm]

        \multicolumn{2}{l}{\tt common/vev/v1,v2} \\

        \hskip 9mm {\tt v1,v2} & Higgs vacuum expectation values $v_1,
        v_2$\\[2mm]

        \multicolumn{2}{l}{\tt common/yukawa/yl(3),yu(3),yd(3)} \\

        \hskip 9mm {\tt yl(3)} & charged lepton Yukawa couplings $Y_e$,
        $Y_\mu$, $Y_\tau$ (complex)\\

        \hskip 9mm {\tt yu(3)} & Running $\overline{\mathrm{MS}}$ up-quark
        Yukawa couplings at $m_t$ scale: $Y_u$, $Y_c$, $Y_t$ \\

        \hskip 9mm {\tt yd(3)} & Running $\overline{\mathrm{MS}}$ down-quark
        Yukawa couplings at $m_t$ scale: $Y_u,Y_c,Y_t$ \\[2mm]

        \multicolumn{2}{l}{\tt common/gmass/gm3,gm2,gm1} \\

        \hskip 9mm {\tt gm1,gm2} & $U(1),SU(2)$ gaugino masses $M_1,M_2$
        (complex)\\

        \hskip 9mm {\tt gm3} & $SU(3)$ gaugino mass $M_3$\\[2mm]

        \multicolumn{2}{l}{\tt
          common/msoft/lms(3,3),rms(3,3),ums(3,3),dms(3,3),qms(3,3)}\\

        \hskip 9mm {\tt lms(3,3),rms(3,3)} & hermitian slepton soft mass
        matrices $m_L^2$, $m_E^2$ (complex)\\

        \hskip 9mm {\tt ums(3,3),dms(3,3),qms(3,3)} & hermitian squark soft
        mass matrices $m_U^2$, $m_D^2$, $m_Q^2$ (complex)\\[2mm]

        \multicolumn{2}{l}{\tt
          common/soft/ls(3,3),ks(3,3),ds(3,3),es(3,3),us(3,3),ws(3,3)}\\

        \hskip 9mm {\tt ls(3,3),ds(3,3),us(3,3)} & trilinear soft SUSY breaking terms $A_l$, $A_d$, $A_u$ (complex) \\

        \hskip 9mm {\tt ks(3,3),es(3,3),ws(3,3)} & trilinear
        ``non-holomorphic'' soft SUSY breaking terms $A'_l$, $A'_d$, $A'_u$
        (complex) \\
      
      \end{tabular}
    \end{center}
  }
  \caption{Common blocks storing the MSSM Lagrangian parameters. }
  \label{tab:mssmpar}
\end{table}

\begin{table}[htbp]
  {\small \begin{center} \begin{tabular}{lp{1mm}p{85mm}}
        Common block and variables && Masses and mixing matrices \\[2mm]
 
        \multicolumn{3}{l}{\tt common/fmass/em(3),um(3),dm(3)} \\

        \hskip 9mm {\tt em(3)} && Charged lepton pole masses $m_e,m_\mu,m_\tau$ \\

        \hskip 9mm {\tt um(3)} && Running $\overline{\mathrm{MS}}$ up-quark
        masses at the $m_t$ scale: $m_u,m_c,m_t$ \\

        \hskip 9mm {\tt dm(3)} && Running $\overline{\mathrm{MS}}$ down-quark
        masses at the $m_t$ scale: $m_u,m_c,m_t$ \\[2mm]

        \multicolumn{3}{l}{\tt common/hmass/cm(2),rm(2),pm(2),zr(2,2),zh(2,2)} \\

        \hskip 9mm {\tt rm(2)} && neutral CP-even Higgs masses ${\tt rm(1)} =
        M_H$, ${\tt rm(2)} = M_h$ \\

        \hskip 9mm {\tt pm(2)} && neutral CP-odd Higgs mass {\tt pm(1)} and
        Goldstone mass {\tt pm(2)} \\

        \hskip 9mm {\tt cm(2)} && charged Higgs mass {\tt cm(1)} and charged
        Goldstone mass {\tt cm(2)} \\

        \hskip 9mm {\tt zr(2,2)} && CP-even Higgs mixing matrix $Z_R$ \\

        \hskip 9mm {\tt zh(2,2)} && CP-odd and charged Higgs mixing matrix
        $Z_H$ \\[2mm]

        \multicolumn{3}{l}{\tt common/charg/fcm(2),zpos(2,2),zneg(2,2)} \\

        \hskip 9mm {\tt fcm(2)} && chargino masses $M_{\chi^+_i}$, $i=1,2$ \\

        \hskip 9mm {\tt zpos(2,2),zneg(2,2)} && chargino mixing matrices $Z_+,
        Z_-$ (complex) \\[2mm]

        \multicolumn{3}{l}{\tt common/neut/fnm(4),zn(4,4)}\\

        \hskip 9mm {\tt fnm(4)} && neutralino masses $M_{\chi^0_i}$,
        $i=1\ldots 4$ \\

        \hskip 9mm {\tt zn(4,4)} && neutralino mixing matrix $Z_N$ (complex)
        \\[2mm]

        \multicolumn{3}{l}{\tt common/slmass/vm(3),slm(6),zv(3,3),zl(6,6)} \\

        \hskip 9mm {\tt vm(3)} && sneutrino masses $M_{\tilde\nu_I}, I=1\ldots
        3$ \\

        \hskip 9mm {\tt slm(6)} && charged slepton masses $M_{L_i}, i=1\ldots
        6$ \\

        \hskip 9mm {\tt zv(3,3)} && sneutrino mixing matrix $Z_{\tilde\nu}$
        (complex)\\

        \hskip 9mm {\tt zl(6,6)} && charged slepton mixing matrix $Z_L$
        (complex) \\[2mm]

        \multicolumn{3}{l}{\tt common/sqmass/sum(6),sdm(6),zu(6,6),zd(6,6)} \\

        \hskip 9mm {\tt sum(6)} && up-squark masses $M_{U_i}, i=1\ldots 6$ \\

        \hskip 9mm {\tt sdm(6)} && down-squark masses $M_{D_i}, i=1\ldots 6$ \\
      
        \hskip 9mm {\tt zu(6,6)} && up-squark mixing matrix $Z_U$ (complex) \\
      
        \hskip 9mm {\tt zd(6,6)} && down-squark mixing matrix $Z_D$ (complex) \\
      
      \end{tabular}
    \end{center}
  }
  \caption{Common blocks storing particle masses and mixing matrices.}
  \label{tab:eigenmass}
\end{table}

\section{Resummation of chirally enhanced corrections}
\label{sec:ren}

The resummation of the chirally enhanced corrections, including the
threshold corrections to Yukawa couplings and CKM matrix elements, is
an important new feature added to \code{} from version 2.0.  Such
corrections arise in the case of large values of $\tan\beta$ or large
trilinear SUSY-breaking terms.  They formally go beyond the 1-loop
approximation, but should be included due to their numerical
importance\footnote{It is even possible that the light fermion masses
  and off-diagonal CKM elements are generated entirely by
  chirally-enhanced self-energies involving the trilinear $A$-terms
  \cite{radiativemasses}}.  Implementation of the resummation in
\code{} follows the systematic approach of ref.~\cite{JRCRIV} and
takes into accounts all contributions involving sfermions and gauginos
(gluino, chargino and neutralino exchanges).  The level of resummation
is a user selectable option and can be done using the following
routine:\\[2mm] {\small
  \begin{tabular}{ll}

    Routine: & {\tt subroutine set\_resummation\_level(ilev,ierr)} \\[1mm]

    Input: & {\tt ilev=0}:  no resummation\\
   
    & {\tt ilev=1}: \begin{minipage}[t]{100mm} analytical solution
      used for bare Yukawa couplings and the bare CKM matrix elements (i.e. 
      parameters of the superpotential), valid in the ``decoupling limit'' 
      $M_{SUSY}\gg v_1,v_2$ \end{minipage}\\

    & {\tt ilev=2}: \begin{minipage}[t]{100mm} exact iterative numerical
      solution for the bare Yukawa couplings and the bare CKM matrix
      elements.  \end{minipage}\\

    Output: & {\tt ierr=0}: resummation successful\\

    & {\tt ierr<0}: \begin{minipage}[t]{100mm}exact resummation
      ({\tt ilev=2}) requested but failed (no convergence), instead analytical
      resummation in the decoupling limit performed
      successfully \end{minipage}\\

    & {\tt ierr>0}: \begin{minipage}[t]{100mm}resummation failed (both
      for {\tt ilev=1,2}), only 1-loop expressions will be used in
      calculations of the physical observables \end{minipage}\\ \\

    Details of calculations: & Ref.~\cite{JRCRIV}\\
\end{tabular}
} \\
After call to {\tt set\_resummation\_level} with {\tt ilev$\neq$ 0},
\code{} calculates the values of bare Yukawa couplings and CKM matrix
elements (i.e. the values of the MSSM Lagrangian parameters) and
starts to use in loop calculations appropriately corrected effective
Higgs boson and supersymmetric particle couplings, automatically
taking into account resummation of enhanced higher order terms.

One should keep in mind that if the chirally enhanced corrections are
very large relation between bare and effective physical fermion masses
and CKM matrix elements involve a significant degree of fine-tuning
and one might also encounter numerical instabilities using the
program.  Therefore, the routines performing the resummation should be
used with care.  One can reasonably assume that resummation works
properly in the decoupling limit $v_1,v_2\ll M_{SUSY}$ as long as the
difference between the bare and physical quantities is at least not
significantly larger than the physical values themselves.  Setting the
actual ``safety condition'' is left to the \code{} users.  To
facilitate that, the blocks {\tt SFLAV\_CHIRAL\_YUKAWA} and {\tt
  SFLAV\_CHIRAL\_CKM} in the \code{} output file list the relative
size of differences between the bare Yukawa couplings and CKM matrix
elements of the superpotential and the (effective) physical
quantities, calculated as
\bea
\delta X_{corr} = \left|{X_{\rm{bare}} - X_{\rm{effective}}\over
    X_{\rm{effective}}}\right|
\eea

One can use such output to define conditions rejecting points of the
MSSM parameters space where the resummation effects are too large and
calculations cannot be trusted.  In our numerical experience, the
stability of \code{} results requires the relative size of the
resummed loop corrections to be at most of order one for CKM elements
and Yukawa couplings of 2nd and 3rd generation.  Thus, if the chosen
input respects 't Hooft's naturalness argument \cite{Crivellin:2008mq,
  Crivellin:2010gw}, also the resummation of all chirally enhanced
effects can be performed analytically in the decoupling limit and is
stable numerically.

\section{List of processes}
\label{sec:proc}

In this section we list the set observables whose computation is
implemented in \code{} v2.5.  QCD corrections and hadronic matrix
elements are extracted mostly from various analyses done within the
Standard Model.  They are assumed to work reasonably well also in the
MSSM since supersymmetric strong corrections from gluino and squarks
are suppressed by large masses of these particles.

The values of the hadronic matrix elements are calculated using the
lattice QCD techniques and thus carry significant theoretical
uncertainties.  Therefore, in \code, hadronic matrix element estimates
and other QCD related quantities are treated as external parameters.
They are initialized to the default values listed below for each
observable and can be directly modified by the users by changing the
relevant variables in the common blocks where they are stored, or
simpler, modifying entries of the {\t Block SFLAV\_HADRON} in the file
{\tt susy\_flavor.in}. Currently most of the hadronic (and related)
input parameters used in \code{} are taken from the Table~3 of
Ref.~\cite{AB}.

In most cases, QCD and hadronic corrections are known to a precision
at the level of few percent to tens of percent, while variations of
supersymmetric flavor and CP-violating parameters can change
observables by orders-of-magnitude.  Thus, as long as the MSSM
parameters are not measured very precisely, the current implementation
of strong corrections is sufficient for analyses performed in the
framework of the general MSSM.

Although \code{} is designed to calculate flavor-related observables,
it is convenient to evaluate within one code also the CP-even Higgs
mass $m_h$, often used as a constraint on the MSSM parameters.
Therefore, in \code{} v2.5 we added the routine calculating the
approximate 2-loop estimate of the neutral CP-even Higgs mass $m_h$,
based on Refs.~\cite{Heinemeyer:1999be, Haber:1996fp}. For precise
calculations of this mass other public SUSY generators should be used.

\subsection{$g-2$ magnetic moment anomaly for leptons}
\label{sec:gminus2}

Anomalous magnetic moment of leptons are defined as the coefficient
$a_{l^I}\equiv (g_I-2)/2$ in the effective Hamiltonian for the
flavor-diagonal lepton-lepton-photon interaction:
\bea \label{eq:gminus2}
{\cal H}_l = - {e\over 4 m_{l^I}} a_I \bar l^I \sigma_{\mu\nu} l^I
  F^{\mu\nu}\;,
\eea
where $I=1,2,3$ is the generation index of the lepton\footnote{The
  measurement of the anomalous magnetic moment of the muon is used to
  determine $\alpha$.  In order to consider the possible effect of new
  physics one needs an independent determination of $\alpha$
  \cite{Girrbach:2009uy} - e.g.  one can use the
  measurements of the Rubidium atom \cite{alpha}.}.   In \code{}
supersymmetric contribution to $(g-2)$ anomaly (to be added to the SM
one) is calculated by the routine:\\[2mm] {\small
\begin{tabular}{lp{5mm}p{100mm}}

Routine: && {\tt double precision function g\_minus\_2\_susy(I)} \\

Input: && $I=1,2,3$ for $e,\mu,\tau$ respectively\\

Output: && SUSY contribution to $a_I=(g_I-2)/2$ for the charged lepton
specified by $I$\\

QCD related factors: && none, QCD corrections are small and not
included\\

Details of calculations: && Performed by authors, unpublished\\

\end{tabular}
}

\subsection{Electric Dipole Moments of charged leptons}
\label{sec:edml}

Lepton EDMs are defined as another coefficient $d_{l^I}$ in the
effective Hamiltonian for the flavor-diagonal lepton-lepton-photon
interaction:
\bea \label{eq:edm}
{\cal H}_l = {i d_{l^I}\over 2}\bar l^I \sigma_{\mu\nu} \gamma_5 l^I
F^{\mu\nu}\;,
\eea
where $I=1,2,3$ is again the generation index of the lepton.   In
\code{} lepton EDM is calculated by:\\[2mm]
{\small
\begin{tabular}{lp{5mm}p{103mm}}

Routine: && {\tt double precision function edm\_l(I)} \\

Input: && $I=1,2,3$ for $e,\mu,\tau$ respectively\\

Output: && EDM for the charged lepton specified by $I$ (in the units
$e~cm$)\\

QCD related factors: && none, QCD corrections are small and not
included\\

Details of calculations: && Ref.~\cite{POROSA} (note that EDM are
defined there with opposite relative sign to \code{} convention)\\

\end{tabular}
}

\subsection{Neutron Electric Dipole Moment}
\label{sec:edmn}

The neutron EDM can be approximated by the sum of the electric dipole
moments of the constituent quarks plus contributions from the
chromoelectric dipole moments (CDM) of quarks and gluons.  The EDMs of
the individual quarks are defined analogously to eq.~(\ref{eq:edm}).
The CDM $c_q$ of the quark $q$ is defined as:
\bea
{\cal H}_c = - \frac{i c_q}{2} \bar{q} \sigma_{\mu\nu} \gamma_5 T^a q
G^{\mu\nu a}.
\label{eq:cdmdef}
\eea
The gluonic dipole moment $c_g$ is defined as:
\bea
{\cal H}_g = - \frac{c_g}{6} f_{abc} G^a_{\mu\rho} G^{b\rho}_{\nu}
G^c_{\lambda\sigma}\epsilon^{\mu\nu\lambda\sigma}.
\label{eq:gdmdef}
\eea
The exact calculation of the neutron EDM requires knowledge of its
hadronic wave function.  \code{} uses the formulae:
\bea
E_n = \eta_{ed} d_d +\eta_{eu} d_u + e(\eta_{cd} c_d + \eta_{cu} c_u)
+\frac{e\eta_g\Lambda_X}{4\pi}c_g
\label{eq:fullneut}
\eea
where $\eta_i$ and $\Lambda_X$ are QCD correction
factors~\cite{ETAQCD} and the chiral symmetry breaking
scale~\cite{QUARKMODEL}, respectively.  Various models give
significantly different factors $\eta_i$.  Thus the \code{} result
should be treated as an order of magnitude estimate only.  The
calculations are performed by calling\\[2mm] {\small
\begin{tabular}{lp{5mm}p{95mm}}

Routine && {\tt double precision function edm\_n()} \\

Input && none\\

Output && neutron EDM\\

QCD related factors: \\

\multicolumn{3}{l}
{\tt common/edm\_qcd/eta\_ed,eta\_eu,eta\_cd,eta\_cu,eta\_g,alamx} \\

\hskip 9mm $\eta_{ed}$ && \hskip 9mm ${\tt eta\_ed} = 0.79 $\\

\hskip 9mm $\eta_{eu}$ && \hskip 9mm ${\tt eta\_eu} = -0.2 $\\

\hskip 9mm $\eta_{cd}$ && \hskip 9mm ${\tt eta\_cd} = 0.59 $\\

\hskip 9mm $\eta_{cu}$ && \hskip 9mm ${\tt eta\_cu} = 0.3 $\\

\hskip 9mm $\eta_{g}$ && \hskip 9mm ${\tt eta\_g} = 3.4 $\\

\hskip 9mm $\Lambda_{X}$ && \hskip 9mm ${\tt alamx} = 1.18 $\\

Details of calculations: && Ref.~\cite{POROSA}\\

\end{tabular}
}

\subsection{$\mu\ra e\gamma$ and $\tau\ra e\gamma,\mu\gamma$
decay rates}
\label{sec:llgamma}

The branching ratios for the flavor violating decays of a heavy lepton
into a lighter lepton and photon are given by:
\bea\label{eq:llgamma}
Br(l^J\to l^I\gamma) = {48\pi^2 e^2Br(l^J\to e\bar\nu \nu)\over
  m_{l^J}^2 G_F^2} \left(|C_L^{JI}|^2 + |C_R^{JI}|^2\right)\;,
\eea
where $C_{L,R}^{IJ}$ are the relevant Wilson coefficients calculated
from the 1-loop lepton-photon triangle diagram with an on-shell
photon.  The branching ratios are calculated by\\[2mm] {\small
\begin{tabular}{lp{5mm}p{103mm}}

Routine: && {\tt double precision function br\_llg(J,I)} \\

Input: && $J,I=1,2,3$ for $e,\mu,\tau$ respectively\\

Output: && branching ratios for $\mu\to e \gamma$ decay ($J=2,I=1$)
and $\tau\to e\gamma, \mu\gamma$ decays ($J=3,I=1,2$)\\

QCD related factors: && none, QCD corrections are small and not
included\\

Details of calculations: && Performed by authors, unpublished\\

\end{tabular}
}

\subsection{$K^0_L\ra \pi^0 \bar\nu \nu$ and $K^+\ra \pi^+ \bar\nu \nu$
decay rates}
\label{sec:kpivv}

The relevant part of the effective Hamiltonian generated by the top
quark and SUSY particle exchanges can be written as
\begin{equation}\label{eq:kpivvham} 
{\cal H}_{\rm eff}={G_{\rm F} \over{\sqrt 2}}{\alpha\over 2\pi
  \sin^2\theta_{\rm w}} \sum_{l=e,\mu,\tau} \left[X_L (\bar
  sd)_{V-A}(\bar\nu_l\nu_l)_{V-A}+ X_R (\bar
  sd)_{V+A}(\bar\nu_l\nu_l)_{V-A}\right].
\end{equation}
The branching ratios for the $K\ra \pi \nu \bar \nu$ decays are then
given by
\bea
\label{bkpnZ}
{Br}(K^+\ra \pi^+ \bar\nu \nu) = \kappa_+ \left[ \left({ \imag
      (X_L+X_R) \over \lambda^5} \right)^2 + \left( {\real
      (K^{\star}_{cs}K_{cd}) \over \lambda} P_c + {\real (X_L + X_R)
      \over\lambda^5} \right)^2 \right]
\eea
\bea
\label{bklpnZ}
\mathrm{Br}(K^0_L\ra \pi^0 \bar\nu \nu)=\kappa_L \left({\imag (X_L +
    X_R) \over \lambda^5}\right)^2
\eea
where $\kappa$~\cite{Mescia:2007kn}, $\lambda$ (the Wolfenstein
parameters~\cite{WO}), and the NLO charm quark contribution
$P_c$~\cite{BB98,BB3,GORBAHN} can be modified by \code{} users (note
that $\kappa$ and $P_c$ depend on $V_{us}$, $m_c$ and $\alpha_s$)
%
%
%
%
%
The calculations of the branching ratios are performed by calling \\[2mm]
{\small 
\begin{tabular}{lp{5mm}p{95mm}}
Routine && {\tt subroutine k\_pivv(br\_k0,br\_kp)} \\
Input && none\\
Output && ${\tt br\_k0} = Br(K^0_L\ra \pi^0 \bar\nu \nu)$ \\
       && ${\tt br\_kp} = Br(K^+\ra \pi^+ \bar\nu \nu)$\\
QCD related factors \\
\multicolumn{3}{l}
{\tt common/kpivv/ak0,del\_ak0,akp,del\_akp,pc,del\_pc,alam} \\
\hskip 9mm $\kappa_L \pm \Delta\kappa_L$ && \hskip 9mm ${\tt
  ak0}=2.231\cdot 10^{-10}$, ${\tt del\_ak0}=0.013\cdot 10^{-10}$ \\
\hskip 9mm $\kappa_+ \pm \Delta\kappa_+$ && \hskip 9mm ${\tt
  akp}=5.173\cdot 10^{-11}$, ${\tt del\_akp}=0.025\cdot 10^{-11}$\\
\hskip 9mm $P_c \pm \Delta P_c$ && \hskip 9mm ${\tt pc}=0.41$, ${\tt
  del\_pc}=0.03$\\
\hskip 9mm $\lambda$ && \hskip 9mm ${\tt alam}=0.225$\\
Details of calculations: && Ref.~\cite{BEJR}\\
\end{tabular}
}

\subsection{$B_d^0\ra l^{I+} l^{J-}$ and $B_s^0\ra l^{I+} l^{J-}$
  decay rates}
\label{sec:bll}

The general expression for these branching ratios are rather
complicated and can be found in~\cite{DRT}\footnote{Note that only the
  1-loop electroweak/SUSY contributions to $B_{d,s}^0\ra l^{I+}
  l^{J-}$ are implemented in \code{} v2.5. Thus, in the limit of heavy
  SUSY masses \code{} reproduces older SM 1-loop estimates for such
  decays, somewhat higher that the NLO result given recently for
  $B_s\to\mu^+\mu^-$ in~\cite{BGGI}}.  For most users it is sufficient
to know that, in addition to the MSSM parameters, the dilepton $B$
decays depend on the $B$ meson masses and the hadronic matrix elements
of the down quark vector and scalar currents:
\bea
\bra{0}\overline{b}\gamma_{\mu} P_{L(R)} s \ket{B_{s(d)}(p)} \ &=&
\ -(+) \frac{i}{2} p_{\mu} f_{B_{s(d)}} \;, \label{np1} \\[3mm]
\bra{0}\overline{b} P_{L(R)} s \ket{B_{s(d)}(p)} \ &=& \ +(-)
\frac{i}{2} \, \frac{M_{B_{s(d)}}^{2} f_{B_{s}}}{m_{b}+m_{s(d)}}
\;,  \label{np2}
\eea
where $p_{\mu}$ is the momentum of the decaying $B_{s(d)}$-meson of
mass $M_{B_{s(d)}}$.  The $B_d^0\ra l^{I+} l^{J-}$ and $B_s^0\ra
l^{I+} l^{J-}$ decay branching ratios are calculated by:\\[2mm]
{\small
\begin{tabular}{lp{5mm}p{105mm}}

Routine && {\tt double precision function b\_ll(K,L,I,J)} \\

Input && $I,J=1,2,3$ - outgoing leptons generation indices \\
 
&& $K,L$ - generation indices of the valence quarks of the $B^0$
meson: setting $(K,L) = (3,1), (1,3), (3,2)$ and $(2,3)$ chooses
respectively $B_d^0$, $\bar B_d^0$, $B_s^0$ and $\bar B_s^0$ decay \\

Output && Branching ratios of the decay defined by $K,L,I,J$\\

QCD related factors \\

\multicolumn{3}{l}
{\tt
  common/meson\_data/dmk,amk,epsk,fk,dmd,amd,fd,amb(2),dmb(2),gam\_b(2),fb(2)}
\\

\hskip 9mm $M_{B_d}$ && \hskip 9mm ${\tt amb(1)}=5.2794$ \\

\hskip 9mm $M_{B_s}$ && \hskip 9mm ${\tt amb(2)}=5.368$ \\

\hskip 9mm $f_{B_d}$ && \hskip 9mm ${\tt fb(1)}=0.193$ \\

\hskip 9mm $f_{B_s}$ && \hskip 9mm ${\tt fb(2)}=0.232$ \\

Details of calculations: && Ref.~\cite{DRT}\\

\end{tabular}
}

\subsection{$B\to (D) \tau \nu$ decay rates}
\label{sec:bdtaunu}

\code{} calculates $Br(B\to \tau \nu)$ $Br(B\to D \tau \nu)$ and
$Br(B\to D^\star \tau \nu)$ including the SM and the charged Higgs
contribution.  The chirally enhanced corrections to Yukawa couplings
from SUSY sectors, which also affect the charged Higgs contribution,
are included.  The relevant part effective Hamiltonian reads as:
\bea
H_{eff}^I = {4G_F V_{qb} \over \sqrt{2}} \left[ \left( \bar q
  \gamma_\mu P_L b \right) \left(\bar \tau \gamma_\mu P_L \nu \right)
  + C^L_q \left( \bar q P_L b \right) \left(\bar \tau P_L \nu \right)
  + C^R_q \left( \bar q P_R b \right) \left(\bar \tau P_L \nu \right)
  \right]\;,
\eea
where $q=u$ for $B\to\tau\nu$ and $q=c$ for $B\to D(D^\star)\tau\nu$
decays. The New Physics $C_q^{L(R)}$ contributions come from the
modification of the effective Yukawa couplings and read as
\bea
C^{L}_q &\approx& - {\sqrt{2} \over 4 m_{H^+}^2 G_F V_{qb} }
\Gamma_{qb}^{H^+RL}\Gamma_{\nu\tau}^{H^+LR\star}\;,\\
C^{R}_q &\approx& - {\sqrt{2} \over 4 m_{H^+}^2 G_F V_{qb} }
\Gamma_{qb}^{H^+LR}\Gamma_{\nu\tau}^{H^+LR\star}\;,
\eea
with
$\Gamma_{qb}^{H^+LR},\Gamma_{qb}^{H^+RL},\Gamma_{\nu\tau}^{H^+LR}$
defined in eqs.~(48), (50) of ref.~\cite{JRCRIV}.

The decay rates are given by~\cite{BTAUNUFORMULAE}:
\bea
Br(B\to \tau \nu) &=& {G_F^2|V_{ub}|^2\over 8\pi}m_\tau^2m_B
f_B^2\tau_B \left(1 - \frac{m_\tau^2}{m_B^2} \right)^2 \left|1 +
\frac{m_B^2}{m_b m_\tau} (C_u^R - C_u^L) \right|^2 \\
{Br(B\to D\tau \nu) \over Br(B\to D l \nu)} &=& R_D \left( 1 + 1.5
\; \mathrm{Re} (C^R_c + C_c^L) + 1.0 \left| C^R_c + C_c^L \right|^2
\right)\\
{Br(B\to D^\star\tau \nu) \over Br(B\to D^\star l \nu)} &=&
R_{D^\star}\left( 1 + 0.12\; \mathrm{Re} (C^R_c - C_c^L) + 0.05 \left|
C^R_c - C_c^L \right|^2 \right)
\eea
where $R_D$ and $R_{D^\star}$ are the respective ratios calculated
within the SM.

Branching ratios are calculated by:\\[2mm] {\small
\begin{tabular}{lp{5mm}p{105mm}}

Routine && {\tt subroutine b\_taunu(br\_taunu,br\_dtaunu,br\_dstaunu)} \\

Input && none \\
 
Output && ${\tt br\_taunu} = Br(B^+\to \tau^+ \nu)$ \\[2mm]

&& ${\tt br\_dtaunu} = {Br(B\to D\tau \nu) \over Br(B\to D l \nu)}$\\[2mm]

&& ${\tt br\_dstaunu} = {Br(B\to D^\star\tau \nu) \over Br(B\to
  D^\star l \nu)}$\\[2mm]

QCD related factors \\

\multicolumn{3}{l}
{\tt
  common/meson\_data/dmk,amk,epsk,fk,dmd,amd,fd,amb(2),dmb(2),gam\_b(2),fb(2)}
\\

\hskip 9mm $f_{B_d}$ && \hskip 9mm ${\tt fb(1)}=0.193$ \\

\multicolumn{3}{l}
{\tt
  common/dtau\_data/dmbp, rd, del\_rd, rds, del\_rds}
\\

\hskip 9mm $M_{B+_u}$ && \hskip 9mm ${\tt dmbp}=5.27917$ \\
\hskip 9mm $R_D\pm \Delta R_D$ && \hskip 9mm ${\tt rd}=0.297$, ${\tt
  del\_rd}=0.017$ \\
\hskip 9mm $R_{D^\star}\pm \Delta R_{D^\star}$ && \hskip 9mm ${\tt
  rds}=0.252$, ${\tt del\_rds}=0.003$ \\

Details of calculations: && Ref.~\cite{BTAUNUFORMULAE, JRCRIV}\\

\end{tabular}
}

\subsection{$B^0\ra X_s \gamma$ decay rate}
\label{sec:bsgamma}

Both the SUSY contributions and the QCD corrections to the calculation
of the $B^0\ra X_s \gamma$ decay rate are quite complex.   Their
implementation in \code{} is based on the SUSY loop calculations
performed by the authors (not published in a general form) and on the
QCD evolution published in~\cite{CMM98}.   There are no user-accessible
QCD factors apart from the arguments of the {\tt bxg\_nl}
routine.\\[2mm]
{
\begin{tabular}{lp{5mm}p{95mm}}

Routine && {\tt double precision function bxg\_nl(del,amiu\_b)}\\

Input && {\tt del} - relative photon energy infrared cutoff scale,
$E_\gamma\ge (1- {\tt del}) E_\gamma^{max}$, $0 < {\tt del} < 1$\\

 && {\tt amiu\_b} - renormalization scale \\

Output && $Br(B\ra X_s \gamma)$.\\

Details of calculations: && General SUSY diagrams unpublished, QCD
corrections based on~\cite{CMM98}\\

\end{tabular}
}

\subsection{$t\to ch,uh$  decay rates}
\label{sec:tqh}

In \code{} v2.5 rare decays of the top quark to a CP-even Higgs boson
and lighter up-type quarks, $t\ra ch,uh$, has also been included based
on Ref.~\cite{T2UH}.  The expression for the relevant branching ratio
is given by
\begin{eqnarray}
Br(t\to q h ) = {m_t \left( 1 - \frac{m_{h}^{2}}{m_{t}^{2}}
  \right)^{2} \over 32\pi \Gamma_{t\to bW}} \left[ 1.018 \left(
  |C^{(h)}_{L}|^{2} + |C^{(h)}_{R}|^{2} \right) + \frac{0.098
    m_{t}^2}{v}\ \Re e \left( C_{R}^{(h)\star} C_{R}^{(g)} +
  C_{L}^{(h)\star} C_{L}^{(g)} \right) \right]
\label{gammaf}
\end{eqnarray}
where $q$ can be either $c$ or $u$, $m_t$ denotes the top quark pole
mass, $v$ is the SM Higgs vev, $C_{L,R}^{(h)}$ are form factors for
the effective flavor violating Higgs-up quark coupling and
$C_{L,R}^{(g)}$ are dipole type form factors for the effective flavor
violating gluon-up quark vertex, all calculated at the scale $\mu =
m_{t}$ (see Ref.~\cite{T2UH} for more details).

Also decays to heavier CP-even Higgs boson, $H$, can be calculated
using this routine, assuming that they are accessible kinematically.
The branching ratios are calculated by\\[2mm] {\small
\begin{tabular}{lp{5mm}p{103mm}}

Routine: && {\tt double precision function br\_suu(I,k)} \\

Input: && $I=1,2$ for, respectively, $u,c$ quark in the final state\\

&& $k=1,2$ for, respectively, $H,h$ Higgs boson in the final state \\

Output: && branching ratios for $t\to ch$ decay ($I=2,k=2$) or $t\to
uh$ decay ($I=1,k=2$) or similar decays to $H$ for $k=1$.\\

QCD related factors: && none\\

Details of calculations: && Ref.~\cite{T2UH}\\

\end{tabular}
}

\subsection{$\bar K^0 K^0$ meson mixing parameters}
\label{sec:kkmix}

\code{} calculates two parameters measuring the amount of CP-violation
in neutral $K$ meson oscillations: $\varepsilon_K$ and the $\bar
K^0-K^0$ mass difference $\Delta M_K$.
\bea
\Delta M_K= 2 \real\langle \bar K^0| H^{\Delta S=2}_{\rm
  eff}|K^0\rangle~,
\eea
\bea
\varepsilon_K=\frac{\exp(i\pi/4)}{\sqrt{2}\Delta M_K} \imag\langle
\bar K^0| H^{\Delta S=2}_{\rm eff}|K^0\rangle~.
\eea
QCD dependent corrections are known with reasonable accuracy for the
$\varepsilon_K$ parameter.   The long distance contributions to $\Delta
M_K$ are large and difficult to control.   Thus the result given by
\code{} for $\Delta M_K$ should be treated as an order of magnitude
estimate only.

Apart from the MSSM parameters, the calculation of the $\bar K^0 K^0$
meson mixing requires knowledge of the meson masses and of the
hadronic matrix elements of the following set of four-quark operators:
\bea 
Q_1^{\rm VLL} &=& (\bar{q}^I_{\alpha} \gamma_{\mu}    P_L q^J_{\alpha})
                  (\bar{q}^I_{\beta} \gamma^{\mu}    P_L q^J_{\beta}),
\nonumber\\ 
Q_1^{\rm LR} &=&  (\bar{q}^I_{\alpha} \gamma_{\mu}    P_L q^J_{\alpha})
                  (\bar{q}^I_{\beta} \gamma^{\mu}    P_R q^J_{\beta}),
\nonumber\\
Q_2^{\rm LR} &=&  (\bar{q}^I_{\alpha}                 P_L q^J_{\alpha})
                  (\bar{q}^I_{\beta}                 P_R q^J_{\beta}),
\nonumber\\
Q_1^{\rm SLL} &=& (\bar{q}^I_{\alpha}                 P_L q^J_{\alpha})
                  (\bar{q}^I_{\beta}                 P_L q^J_{\beta}),
\nonumber\\
Q_2^{\rm SLL} &=& (\bar{q}^I_{\alpha} \sigma_{\mu\nu} P_L q^J_{\alpha})
                  (\bar{q}^i_{\beta} \sigma^{\mu\nu} P_L q^J_{\beta})
\label{eq:4qbase}
\eea
where $\alpha, \beta$ are color indices, for the $\bar{K}^0 K^0$
mixing one should choose flavor indices $I=2$ and $J=1$.   The matrix
elements can be written as:
\bea
\label{eq:4qmatel}
\langle \bar K^0 | Q_1^{\rm VLL}(\mu) | K^0 \rangle &=& \frac{1}{3}
M_K F_K^2 B_1^{\rm VLL} (\mu) ,\nonumber\\
\langle \bar K^0 | Q_1^{\rm LR}(\mu) | K^0 \rangle &=& -\frac{1}{6}
\left( \frac{M_K}{m_s(\mu) + m_d(\mu)} \right)^2 M_K F_K^2 B_1^{\rm
  LR} (\mu) ,\nonumber\\
\langle \bar K^0 | Q_2^{\rm LR}(\mu) | K^0 \rangle &=& \frac{1}{4}
\left( \frac{M_K}{m_s(\mu) + m_d(\mu)} \right)^2 M_K F_K^2 B_2^{\rm
  LR} (\mu) ,\nonumber\\
\langle \bar K^0 | Q_1^{\rm SLL}(\mu) | K^0 \rangle &=& -\frac{5}{24}
\left( \frac{M_K}{m_s(\mu) + m_d(\mu)} \right)^2 M_K F_K^2 B_1^{\rm
  SLL} (\mu) ,\nonumber\\
\langle \bar K^0 | Q_2^{\rm SLL}(\mu) | K^0 \rangle &=& -\frac{1}{2}
\left( \frac{M_K}{m_s(\mu) + m_d(\mu)} \right)^2 M_K F_K^2 B_2^{\rm
  SLL} (\mu),
\eea
where $F_K$ is the $K$-meson decay constant.   By default, \code{} uses
the $B_i^X$ values at the scale $\mu=2$ GeV given in~\cite{BJU} using
the NDR renormalization scheme (quark masses at the scale 2 GeV are
stored in {\tt common/fmass\_high/}).

In addition to the hadronic matrix elements, QCD corrections depend
also on the ``$\eta$'' factors describing the evolution of the
relevant Wilson coefficients from the high to low energy scale.  These
factors are automatically calculated at NLO by \code{}.  For the SM
contribution to the Wilson coefficient of the $Q^{\mathrm{VLL}}$
operator a separate careful calculation of the evolution factors has
been performed~\cite{BJW,HN}.  Therefore \code{} treats this
contribution separately, setting $B_{SM}^{\rm VLL}$ and the
$\eta_{SM}$ factor to default values given in~\cite{GBEPSK}
(see~\cite{BJU} for a very detailed discussion of the structure of the
QCD corrections in $\bar{B}^0B^0$ and $\bar{K}^0K^0$ systems,
including their renormalization scheme dependence and calculations of
the QCD qevolution factors implemented in \code).

The kaon mass difference $\Delta M_K$ and the $\varepsilon_K$
parameter measuring the amount of CP violation in $\bar K^0 K^0$
mixing are calculated by\\[2mm]
{\small
\begin{tabular}{lp{5mm}p{95mm}}

Routine && {\tt subroutine dd\_kaon(eps\_k,delta\_mk)} \\

Input && none\\

Output && ${\tt eps\_k}=\varepsilon_K$ parameter \\

 && ${\tt delta\_mk }=\Delta M_K$ mass difference\\

QCD related factors: \\

\multicolumn{3}{l}
{\tt
  common/meson\_data/dmk,amk,epsk,fk,dmd,amd,fd,amb(2),dmb(2),gam\_b(2),fb(2)}
\\

\hskip 9mm $M_K$ && \hskip 9mm ${\tt amk}=0.497614$ \\

\hskip 9mm Measured $\Delta M_K^{exp}$ && \hskip 9mm ${\tt dmk}=
3.483\cdot 10^{-15}$ \\

\hskip 9mm Measured $\varepsilon_K^{exp}$ && \hskip 9mm ${\tt
  epsk}=2.229\cdot 10^{-3}$ \\

\hskip 9mm $f_K$ && \hskip 9mm ${\tt fk}=0.156$ \\

\multicolumn{3}{l}
{\tt common/bx\_4q/bk(5),bd(5),bb(2,5),amu\_k,amu\_d,amu\_b}\\

\hskip 9mm $B_1^{\rm VLL} (\mu_K)$ && \hskip 9mm ${\tt bk(1)}=0.61$\\

\hskip 9mm $B_1^{\rm SLL} (\mu_K)$ && \hskip 9mm ${\tt bk(2)}=0.76$\\

\hskip 9mm $B_2^{\rm SLL} (\mu_K)$ && \hskip 9mm ${\tt bk(3)}=0.51$\\

\hskip 9mm $B_1^{\rm LR} (\mu_K)$ && \hskip 9mm ${\tt bk(4)}=0.96$\\

\hskip 9mm $B_2^{\rm LR} (\mu_K)$ && \hskip 9mm ${\tt bk(5)}=1.30$\\

\hskip 9mm Renormalization scale $\mu_K$ && \hskip 9mm ${\tt
  amu\_k}=2$\\

\multicolumn{3}{l}
{\tt
  common/sm\_4q/eta\_cc,eta\_ct,eta\_tt,eta\_b,bk\_sm,bd\_sm,bb\_sm(2)}\\

\hskip 9mm $B_{SM}^{\rm VLL}$ && \hskip 9mm ${\tt bk\_sm}=0.724$\\

\hskip 9mm $\eta_{cc}$ && \hskip 9mm ${\tt eta\_cc}=1.86$\\

\hskip 9mm $\eta_{ct}$ && \hskip 9mm ${\tt eta\_ct}=0.496$\\

\hskip 9mm $\eta_{tt}$ && \hskip 9mm ${\tt eta\_tt}=0.577$\\

Details of calculations: && Ref.~\cite{BJU,BCRS}\\

\end{tabular}
}

\subsection{$\bar D^0 D^0$ meson mass difference}
\label{sec:ddmix}

Calculations of the mass difference $\Delta m_D$ of the neutral $D$
mesons have large theoretical uncertainties due to unknown
long-distance strong corrections.  Thus, as in the case of $\Delta
m_K$, the \code{} result for $\Delta m_D$ should be treated as an
order of magnitude estimate only.

The structure of strong corrections is analogous to those in the $K$
meson system.  However, in this case hadronic matrix elements and QCD
evolution calculations available in the literature are much less
refined.  \code{} uses the NLO evolution for the ``$\eta$'' factors
and sets, by default, all the relevant hadronic matrix elements
$B_i=1$, i.e.  it uses the ``vacuum saturation'' approximation (this
can be changed easily when new results become available).\\[2mm]
{\small
\begin{tabular}{lp{5mm}p{95mm}}

Routine && {\tt subroutine uu\_bmeson(delta\_md)} \\

Input && none\\ 

Output && ${\tt delta\_md }=\Delta M_D$ mass difference\\

QCD related factors: \\

\multicolumn{3}{l}
{\tt
  common/meson\_data/dmk,amk,epsk,fk,dmd,amd,fd,amb(2),dmb(2),gam\_b(2),fb(2)}
\\

\hskip 9mm $M_D$ && \hskip 9mm ${\tt amd}=1.8645$ \\

\hskip 9mm Measured $\Delta M_D^{exp}$ && \hskip 9mm ${\tt dmd}=
4.61\cdot 10^{-14}$ \\

\hskip 9mm $f_D$ && \hskip 9mm ${\tt fd}=0.2$ \\

\multicolumn{3}{l}
{\tt common/bx\_4q/bk(5),bd(5),bb(2,5),amu\_k,amu\_d,amu\_b}\\

\hskip 9mm $B_1^{\rm VLL} (\mu_D)$ && \hskip 9mm ${\tt bd(1)}=1$\\

\hskip 9mm $B_1^{\rm SLL} (\mu_D)$ && \hskip 9mm ${\tt bd(2)}=1$\\

\hskip 9mm $B_2^{\rm SLL} (\mu_D)$ && \hskip 9mm ${\tt bd(3)}=1$\\

\hskip 9mm $B_1^{\rm LR} (\mu_D)$ && \hskip 9mm ${\tt bd(4)}=1$\\

\hskip 9mm $B_2^{\rm LR} (\mu_D)$ && \hskip 9mm ${\tt bd(5)}=1$\\

\hskip 9mm Renormalization scale $\mu_D$ && \hskip 9mm ${\tt
  amu\_d}=2$\\

\multicolumn{3}{l}
{\tt
  common/sm\_4q/eta\_cc,eta\_ct,eta\_tt,eta\_b,bk\_sm,bd\_sm,bb\_sm(2)}\\

\hskip 9mm $B_{SM}^{\rm VLL}$ && \hskip 9mm ${\tt bd\_sm}=1$\\

Details of calculations: && Performed by authors, unpublished \\

\end{tabular}
}

\subsection{$\bar B_d^0 B_d^0$ and $\bar B_s^0 B_s^0$ meson mixing
  parameters}
\label{sec:bbmix}

Mixing and CP violation phenomena are also observed in the neutral $B$
meson systems.  In particular, the mass differences in the $\bar B_d^0
B_d^0$ and $\bar B_s^0 B_s^0$ oscillations have been measured,
\bea
\Delta M_{B_{d(s)}} = 2 \left|\langle \bar B^0_{d(s)}| H^{\Delta B=2}_{\rm
  eff}|B^0_{d(s)}\rangle\right|~.
\label{eq:bbmix}
\eea
The time-dependent CP asymmetry in $B_d \to J/\psi K_s$ decays,
$a_{J/\psi K_s}=\sin 2\beta_{eff} \sin \Delta M_{B_d} t$, is also
measured.   It can be related to the argument of the $\Delta F=2$
hadronic matrix element:
\bea
2\beta_{eff} = \mathrm{Arg} \left[\langle \bar B^0_{d}| H^{\Delta
    B=2}_{\rm eff}|B^0_{d(s)}\rangle\right]~.
\label{eq:bbasym}
\eea
As experimental definitions of CP asymmetries are often
convention-dependent, \code{} gives as a more universal output
directly real and imaginary parts of the $\Delta F=2$ matrix element,
which can be further used in various asymmetry calculations.

In addition to the MSSM parameters, theoretical calculations of
$\Delta m_{B_d}$ and $\Delta m_{B_s}$ depend, as for $K$ and $D$
oscillations, on the relevant hadronic matrix elements and QCD
evolution factors.  The formulae for $\bar{B}^0B^0$ mixing can be
obtained by making the obvious replacements in the formulae presented
in Sec.~\ref{sec:kkmix}.  Currently \code{} uses the same set of
$B_i$ factors for both the $B_d$ and $B_s$ sectors, but it leaves the
possibility to distinguish between them in future, if necessary.  For
this one needs to independently initialize the arrays {\tt bb(1,i)}
($B_d$ meson hadronic matrix elements) and {\tt bb(2,i)} ($B_s$ meson
hadronic matrix elements) stored in {\tt common/bx\_4q/}.

The values of the $B$ meson masses and coupling constants are the same
as those listed in Sec.~\ref{sec:bll}.  $\Delta M_{B_{d(s)}}$ is
calculated by:\\[2mm]
{\small
\begin{tabular}{llp{100mm}}

Routine && {\tt subroutine dd\_bmeson(i,delta\_mb,dmb\_re,dmb\_im)} \\

Input && $i=1,2$ - generation index of the lighter valence quark in
the $B^0$ meson, i.e.   $i=2$ chooses $B_s^0$ and $i=1$ chooses
$B_d^0$.\\

Output && ${\tt delta\_mb}=\Delta m_{B_d}(\Delta m_{B_s})$ for
$i=1(2)$\\

&& ${\tt dmb\_re}=\mathrm{Re} [\langle \bar B^0_{d(s)}| H^{\Delta
    B=2}_{\rm eff}|B^0_{d(s)}\rangle]$ for $i=1(2)$\\

&& ${\tt dmb\_im}=\mathrm{Im} [\langle \bar B^0_{d(s)}| H^{\Delta
    B=2}_{\rm eff}|B^0_{d(s)}\rangle]$ for $i=1(2)$\\

QCD related factors:\\

\multicolumn{3}{l} 
{\tt
  common/meson\_data/dmk,amk,epsk,fk,dmd,amd,fd,amb(2),dmb(2),gam\_b(2),fb(2)}
\\

\hskip 9mm Measured $\Delta M_{B_d}^{exp}$ && \hskip 9mm ${\tt
  dmb(1)}= 3.337\cdot 10^{-13}$ \\

\hskip 9mm Measured $\Delta M_{B_s}^{exp}$ && \hskip 9mm ${\tt
  dmb(2)}= 1.17\cdot 10^{-11}$ \\

\hskip 9mm Measured lifetime $\Gamma_{B_d}^{exp}$ && \hskip 9mm ${\tt
  gam\_b(1)}= 1.519\cdot 10^{-12}$ \\

\hskip 9mm Measured lifetime $\Gamma_{B_s}^{exp}$ && \hskip 9mm ${\tt
  gam\_b(1)}= 1.512\cdot 10^{-12}$ \\

\multicolumn{3}{l}
{\tt common/bx\_4q/bk(5),bd(5),bb(2,5),amu\_k,amu\_d,amu\_b}\\

\hskip 9mm $B_1^{\rm VLL} (\mu_B)$ && \hskip 9mm ${\tt bb(1,1)=bb(2,1)}=0.87$\\

\hskip 9mm $B_1^{\rm SLL} (\mu_B)$ && \hskip 9mm ${\tt bb(1,2)=bb(2,2)}=0.8$\\

\hskip 9mm $B_2^{\rm SLL} (\mu_B)$ && \hskip 9mm ${\tt bb(1,3)=bb(2,3)}=0.71$\\

\hskip 9mm $B_1^{\rm LR} (\mu_B)$ && \hskip 9mm ${\tt bb(1,4)=bb(2,4)}=1.71$\\

\hskip 9mm $B_2^{\rm LR} (\mu_B)$ && \hskip 9mm ${\tt bb(1,5)=bb(2,5)}=1.16$\\

\hskip 9mm Renormalization scale $\mu_B$ && \hskip 9mm ${\tt
  amu\_b}=4.6$\\

\multicolumn{3}{l}
{\tt
  common/sm\_4q/eta\_cc,eta\_ct,eta\_tt,eta\_b,bk\_sm,bd\_sm,bb\_sm(2)}\\

\hskip 9mm $B_{SM B_d}^{\rm VLL}$ && \hskip 9mm ${\tt bb\_sm(1)}=1.22$\\
\hskip 9mm $B_{SM B_s}^{\rm VLL}$ && \hskip 9mm ${\tt bb\_sm(2)}=1.22$\\

\hskip 9mm $\eta_b$ && \hskip 9mm ${\tt eta\_b}=0.55$\\

Details of calculations: && Ref.~\cite{BCRS}
\end{tabular}
}

\section{\code{} output}
\label{sec:output}

Starting from v2.10, the \code{} output is written to the file named
{\tt susy\_flavor.out}.  It has a ``SLHA-like'' structure, i.e. it is
split into ``data blocks''. However, these blocks are \code-specific
and do not follow common SLHA2 standards. The output file of \code{}
v2.5 contains the data blocks listed in Table~\ref{tab:output}.

\begin{table}
\begin{tabular}{lp{115mm}}
Block name & Block content \\[2mm]
{\tt SFLAV\_CONTROL} & \code{} control variables and error code status
\\[1mm]
{\tt SFLAV\_MASS} & full mass spectrum of the MSSM particles after
mass matrix diagonalization\\[1mm]
{\tt SFLAV\_CHIRAL\_YUKAWA} & Relative size of resummed chiral
corrections to Yukawa couplings\\[1mm]
{\tt SFLAV\_CHIRAL\_CKM} & Relative size of resummed chiral
corrections to CKM matrix \\[1mm]
{\tt SFLAV\_DELTA\_F0} & Observables related to $\Delta F = 0$
processes (EDM, $g-2$ anomaly)\\[1mm]
{\tt SFLAV\_DELTA\_F1} & Observables related to $\Delta F = 1$
processes ($l\to l'\gamma$, $K\to\pi\bar\nu\nu$, $B^+ \to \tau^+ \nu$,
$B \to D \tau \nu$, $B \to D^\star \tau \nu$, $B \to X_s \gamma$,
$B_{d,s}\to l^+_i l_j^-$, $t\to c h$, $t\to u h$)\\[1mm]
{\tt SFLAV\_DELTA\_F2} & Observables related to $\Delta F = 2$
processes ($\epsilon_K$, $\Delta m_K$, $\Delta m_D$, $\Delta m_{B_d}$,
$\Delta m_{B_s}$)\\
\end{tabular}
\caption{
Block structure of {\tt susy\_flavor.out} file.
\label{tab:output}
}
\end{table}

The first four blocks in {\tt susy\_flavor.out} are included for
control and test purposes. The block {\tt SFLAV\_CONTROL} lists the
state of control variables defining conventions used for input
parameters, in particular the dimension of sfermion flavor violating
parameters. The block {\tt SFLAV\_MASS} contains a full list of MSSM
particle masses - mass eigenstates of sleptons, squarks, neutralinos
and charginos, physical Higgs boson masses (as mentioned earlier the
estimate of $m_h$ is calculated using the approximate 2-loop formulae
based on Refs.~\cite{Heinemeyer:1999be, Haber:1996fp}) and, for
completeness, the pole lepton masses and running quark masses at $m_t$
scale.  The blocks {\tt SFLAV\_CHIRAL\_YUKAWA} and {\tt
  SFLAV\_CHIRAL\_CKM} show the relative difference of bare
vs. physical Yukawa couplings and CKM matrix elements after
resummation of chiral corrections. If they are too large, $\geq {\cal
  O}(1)$, the perturbative loop calculations may not be converging an
the remaining program output cannot be considered to be fully
reliable.

Finally, the entries of the blocks {\tt SFLAV\_DELTA\_F0}, {\tt
  SFLAV\_DELTA\_F1} and {\tt SFLAV\_DELTA\_F2} contain the values of
the flavor and CP-violating observables given in Table~\ref{tab:proc}.

\section{Summary and Outlook}
\label{sec:summary}

We have presented \code{} v2.5, a tool for calculating important
flavor observables in the general $R$-parity conserving MSSM.  Version
2 of \code{} is capable of calculating:
\begin{itemize}
\item Electric dipole moments of the leptons and the neutron.  
\item Supersymmetric contributions to anomalous magnetic moments
  $g-2$ of leptons.
\item Radiative lepton decays ($\mu\to e\gamma$ and $\tau\to
  \mu\gamma, e\gamma$).
\item Rare Kaon decays ($K^0_L\ra \pi^0\bar\nu\nu$ and $K^+\ra \pi^+
  \bar\nu\nu$).
\item Leptonic $B$ decays ($B_{s,d}\ra l^+ l^-$, $B^+\to \tau^+ \nu$,
  $B\to D \tau \nu$ and $B\to D^\star \tau \nu$).
\item Radiative $B$ decays ($B\to\bar X_s \gamma$).
\item Rare decays of the top quark to Higgs boson ($t\to ch,uh$).
\item $\Delta F=2$ processes ($\bar K^0$--$K^0$, $\bar D$--$D$, $\bar
  B_d$--$B_d$ and $\bar B_s$--$B_s$ mixing).
\end{itemize}
All implemented physical observables can be calculated simultaneously
for a given set of MSSM input parameters.  The calculation of the SUSY
tree-level particle spectrum and flavor mixing matrices are performed
exactly, so the code can be used for a completely general pattern of
soft SUSY breaking terms (including complex phases), without
restrictions on the size of the off-diagonal elements in the sfermion
mass matrices.  Program is written in FORTRAN 77 and runs fairly
quickly; it is capable of producing a reasonably wide-range scan over
the MSSM parameters within hours or days on a typical personal
computer.

In code \code{} v2 the resummation of chirally enhanced corrections
(stemming from large values of $\tan\beta$ and/or large trilinear
$A$-terms) has been implemented using the systematic method developed
in~\cite{JRCRIV}.  Such corrections modify the effective couplings of
supersymmetric particles and charged Higgs bosons and generate
enhanced flavor-changing neutral Higgs couplings, the latter giving
significant contributions to various amplitudes coming from
Higgs-penguin type diagrams.  Thus, \code{} is valid for the whole
parameter space of the general $R$-parity conserving MSSM, a unique
feature currently not shared by other publicly available programs
calculating FCNC and CP violation in SUSY models.

Starting from v2.5, \code{} accepts automatically as input most of
output files from other libraries calculating SUSY processes. Only the
parameters relevant for given problem needs to be defined in the input
file, others are initialized using the predefined default values.

Besides complete routines for calculating the physical observables,
\code{} v2 also provides an extensive library of parton-level Green's
functions and Wilson coefficients of many effective quark and lepton
operators (see Table~\ref{tab:green}).  This set actually contains
many more amplitudes than necessary to compute the quantities listed
in Table~\ref{tab:proc}.  These intermediate building blocks can be
used by \code{} users to calculate observables related to additional
processes, beyond those already fully implemented, by dressing
appropriate combinations of available form factors in QCD corrections
and hadronic matrix elements, without repeating tedious SUSY loop
calculations.  For instance, the form factors implemented in \code{}
for the analysis of $B\ra X_s\gamma$ and $B_{d(s)}\to l^+ l^-$
decays~\cite{MIPORO,DRT} are sufficient to also calculate the $B\to K
l^+ l^-$ decay rate.

The \code{} library is an open project.  We want to gradually add more
features in future versions.  In particular, we plan to:
\begin{itemize}
\item add more observables in the $B$-meson system, like the CP
  asymmetries in $B\to X_s \gamma$ decay, observables associated with
  $B\to K l^+ l^-$ decay and others.
\item include more FCNC related quantities in the top sector, in
  particular $t\to q\gamma, qZ$ and $qg$ decay rates.
\item include the effects of massive neutrinos.
\end{itemize}

With the increasing accuracy of experimental data on flavor and CP
violation in rare processes, it may eventually become possible to not
only constrain the MSSM parameters, but also, if significant
deviations from the SM predictions are found, to recover their actual
values.  For that multi-process analysis, such as the one performed by
\code, will be necessary.  Therefore, we hope that \code{} becomes an
important tool that is useful not only to theorists working on MSSM
but also to experimentalists fitting the MSSM onto current and
forthcoming data from the Tevatron, LHC, and $B$-factories.

\subsection*{Acknowledgments}

\noindent The authors thank A.~Buras, T.~Ewerth, L~Hofer, M.~Misiak,
C.~Savoy, {\L}.~S{\l}awianowska, S.~Pokorski, M. Paraskevas and
K. Suxho for collaboration in performing theoretical calculations used
in \code{} and for helping to check and debug some of its sections.
We would also like to thank W.~Altmannshofer, J. Stockel, D.~Straub,
S.~Frank, D.~Guadagnoli, W.~Porod, M.~Wick, J.~Berger and D.~Ghosh for
checking various parts of the \code{} code and reporting bugs or
inconsistencies.

This work has received funding from the EU Seventh Framework Programme
under grant agreement PITN-GA-2009-237920 (2009-2013). A.C.~is
supported by the Swiss National Science Foundation.  The Albert
Einstein Center for Fundamental Physics is supported by the
``Innovations- und Kooperationsprojekt C-13 of the Schweizerische
Universit\"atskonferenz SUK/CRUS''.  The work by J.R. is supported in
part by National Science Center under research grant
DEC-2011/01/M/ST2/02466 (12.2011-12.2014).  The A.D. research has been
co-financed by the European Union (European Social Fund \&\#8211; ESF)
and Greek national funds through the Operational Program ``Education
and Lifelong Learning'' of the National Strategic Reference Framework
(NSRF) - Research Funding Program: THALIS. Investing in the society of
knowledge through the European Social Fund.  S.J. was supported by the
Science and Technology Facilities Council [grant number ST/H004661/1]
and acknowledges support from the NExT institute and SEPnet.  P.T. is
supported by the Paul and Daisy Soros foundation and the U.S. National
Science Foundation through grant PHY-0757868.

\def\theequation{\Alph{section}.\arabic{equation}}
\begin{appendix}

\setcounter{equation}{0}

\parindent 0cm

\newpage

\section{Installation of the program}
\label{app:inst}

The installation and execution of \code{} is very simple.  On Unix or
Linux systems, just follow these steps :

\begin{enumerate}

\item Download the latest version of the code from \webpage{} and
  unpack it.

\item Change directory into {\tt susy\_{flavor}}.

\item Edit {\tt Makefile} and change {\tt F77 = gfortran} and {\tt
  FOPT = -O -fno-automatic -Wall} into your compiler name and options,
  respectively.

\item To use the {\tt susy\_flavor\_file.f} driver, reading input data
  from {\tt susy\_flavor.in} file, type {\tt make sfile} (or simply
  {\tt make}). To use the {\tt susy\_flavor\_prog.f} driver, where
  input data are initialized directly inside the FORTRAN code, type
  {\tt make sprog}.

\item If everything goes through, the code output is written to the
  file {\tt susy\_flavor.out}.

\item To run the code from now on just type {\tt ./sfile} or {\tt
  ./sprog}.

\end{enumerate}

The authors tested \code{} on Linux machines.  With few
straightforward modifications the procedure describe above can be
adapted to install program on other systems.  A sample set of input
parameters and corresponding \code{} output are listed in the
following appendices.

\newpage

\section{Example of the \code{} initialization sequence}
\label{app:code}

Below we list the contents of {\tt susy\_flavor\_file.f} and {\tt
  susy\_flavor\_prog.f}, the driver files for the \code{} library.
They illustrate the correct initialization sequence for all relevant
MSSM parameters (see Sec.~\ref{sec:init}) and how to perform calls to
the routines calculating physical observables (Sec.~\ref{sec:proc}).

Driver program {\tt susy\_flavor\_file.f}, initializing MSSM
parameters from the input file {\tt susy\_flavor.in} is compact and
simple:
\begin{small}
{\tt
\begin{tabbing}

~~~~~~~\=program susy\_flavor\_file~~~~~~~~~~~~ \= \+ \\
      program susy\_flavor\_header\\
      implicit double precision (a-h,o-z)\\
 \- \\
  {\it c} \> \+    choose MSSM sectors to include \\
      ih = 1            \> {\it         ! Higgs + gauge diagrams included}  \\
      ic = 1           \> {\it          ! chargino diagrams included}  \\
      in = 1            \> {\it         ! neutralino diagrams included}  \\
      ig = 1           \> {\it          ! gluino diagrams included}  \\    
      call set\_active\_sector(ih,ic,in,ig) \> {\it ! set control variables} \\
\\

      call sflav\_input(ilev,ierr) \> {\it ! parameters read from susy\_flavor.in}  \\
      if (ierr.ne.0) write(*,*) 'Error in parameter initialization!' \\
      call set\_resummation\_level(ilev,ierr) \> {\it ! resummation of
        chiral corrections} \\
    if (ierr.ne.0) write(*,*)ierr,'Error in chiral corrections resummation!'\\
    call susy\_flavor \> {\it ! main routine calculating physical  observables} \\
      call sflav\_output(ilev,ierr) \> {\it ! output written to susy\_flavor.out} \\
  end  

\end{tabbing}
}
\end{small}

Driver {\tt susy\_flavor\_prog.f} is longer and more complicated as
all parameters has to be specified inside the code.  Using this diver,
flavor violating entries of sfermion mass matrices has to be given as
dimensionless mass insertions.

\begin{small}

{\tt
\begin{tabbing}

  ~~~~~~~\=program susy\_flavor\_prog~~~~~~~~~~~ \= \+ \\
  implicit double precision (a-h,o-z)\\
  dimension sll(3),slr(3),amsq(3),amsu(3),amsd(3)\\
  double complex slmi\_l(3),slmi\_r(3),slmi\_lr(3,3),slmi\_lrp(3,3)\\
  double complex sqmi\_l(3),sdmi\_r(3),sumi\_r(3)\\
  double complex sdmi\_lr(3,3),sumi\_lr(3,3)\\
  double complex sdmi\_lrp(3,3),sumi\_lrp(3,3)\\
  double complex amg,amgg,amue\\
  common/sf\_cont/eps,indx(3,3),iconv\\
  \- \\
  {\it c} \> \+ {\it Input convention choice:}  \\
  iconv = 1 \> {\it !  SLHA2 input conventions} \- \\
  {\it c} \>  {\it iconv = 2} \> {\it !  hep-ph/9511250 input conventions}  \\
  \\
  {\it c} \>  {\it fixes the treatment of enhanced chiral correction resummation}\\
  {\it c} \>  {\it ilev = 0}      \> {\it ! no resummation, SUSY corrections strictly 1-loop}\\
  {\it c} \> \+ {\it ilev = 1}    \> {\it ! resummation using the decoupling limit}\\
  ilev = 2  \> {\it  ! exact iterative solution, may not always converge} \-\\
  \\
  {\it c} \> \+    choose MSSM sectors to include \\
      ih = 1            \> {\it         ! Higgs + gauge diagrams included}  \\
      ic = 1           \> {\it          ! chargino diagrams included}  \\
      in = 1            \> {\it         ! neutralino diagrams included}  \\
      ig = 1           \> {\it          ! gluino diagrams included}  \\    
      call set\_active\_sector(ih,ic,in,ig) \> {\it ! set control variables}\\
\\
      call sflav\_sm      \> {\it           ! initialize auxiliary SM parameters}\- \\
\\
  {\it c} \> \+ {\it SM basic input initialization}\\
  zm0 = 91.1876d0 \> {\it !  M\_Z}\\
  wm0 = 80.398d0 \> {\it !  M\_W}\\
  alpha\_z = 1/127.934d0 \> {\it !  alpha\_em(M\_Z)}\\
  st2\_new = 0.23116d0 \> {\it !  s\_W$^2$(MSBar)}\\
  call vpar\_update(zm0,wm0,alpha\_z,st2\_new) \- \\
  \\
  {\it c} \> \+ {\it CKM matrix initialization}\\
  alam = 0.2258d0           \> {\it ! lambda}\\
  apar = 0.808d0            \> {\it ! A}\\
  rhobar = 0.177d0          \> {\it ! rho bar}\\
  etabar = 0.360d0          \> {\it ! eta bar}\\
  call ckm\_wolf(alam,apar,rhobar,etabar) \- \\
  \\
  {\it c} \> \+ {\it Fermion mass initialization, input: MSbar running quark masses}\\
  alpha\_s = 0.1172d0        \> {\it ! alpha\_s(MZ)}\\
  top\_scale = 163.1d0\\
  top = 163.1d0             \> {\it ! m\_t(top\_scale)}\\
  bot\_scale = 4.18d0\\
  bot = 4.18d0              \> {\it ! m\_b(bot\_scale)}\\
  call init\_fermion\_sector(alpha\_s,top,top\_scale,bot,bot\_scale) \- \\
  \\
  {\it c} \> \+ {\it Higgs sector parameters}\\
  pm    = 200               \> {\it ! M\_A}\\
  tanbe = 4                \> {\it ! tan(beta)}\\
  amue  = (200.d0,100.d0)   \> {\it ! mu}\\
  call init\_higgs\_sector(pm,tanbe,amue,ierr)\\
  if (ierr.ne.0) stop 'negative tree level Higgs mass$^2$?' \- \\
  \\
  {\it c} \> {\it Gaugino sector parameters. CAUTION: if M1 is set to 0 here then}\\
  {\it c} \> \+ {\it program sets M1 and M2 GUT-related, i.e. M1 = 5/3 s\_W$^2$/c\_W$^2$*M2}\\
\\
  amgg  = (200.d0,0.d0)   \> {\it ! M1 (bino mass)}\\
  amg   = (300.d0,0.d0)   \> {\it ! M2 (wino mass)}\\
  amglu = 600               \> {\it ! M3 (gluino mass)}\\
  call init\_ino\_sector(amgg,amg,amglu,amue,tanbe,ierr)\\
  if (ierr.ne.0) write(*,*) '-ino mass below M\_Z/2?'\- \\
  \\
  {\it c} \> \+ {\it Slepton diagonal soft breaking parameters}\\
  sll(1) = 300.d0           \> {\it ! left selectron mass scale}\\
  sll(2) = 300.d0           \> {\it ! left smuon mass scale}\\
  sll(3) = 300.d0           \> {\it ! left stau mass scale}\\
  slr(1) = 300.d0           \> {\it ! right selectron mass scale}\\
  slr(2) = 300.d0           \> {\it ! right smuon mass scale}\\
  slr(3) = 300.d0           \> {\it ! right stau mass scale} \- \\
  {\it c} \> {\it Slepton LL and RR mass insertions (hermitian matrices)}\\
  {\it c} \> \+ {\it slmi\_x(1),slmi\_x(2), slmi\_x(3) are 12,23,31 entry respectively}\\
  do i=1,3\\
  ~~ slmi\_l(i) = dcmplx(0.d0,0.d0) \> {\it ! slepton LL mass insertion}\\
  ~~ slmi\_r(i) = dcmplx(0.d0,0.d0) \> {\it ! slepton RR mass insertion}\\
  end do \\
  slmi\_l(1) = (2.d-2,3.d-2) \> {\it ! example, non-vanishing LL 12 entry} \- \\
  {\it c} \>  {\it Slepton LR mass insertions, non-hermitian in general}\\
  {\it c} \> \+ {\it All entries dimensionless (normalized to diagonal masses)}\\
  do i=1,3\\
  ~~do j=1,3 \- \\
  {\it c} \> \+ {\it holomorphic LR mixing terms}\\
  ~~~~    slmi\_lr(i,j) = (0.d0,0.d0) \- \\
  {\it c} \> \+ {\it non-holomorphic LR mixing terms}\\
  ~~~~     slmi\_lrp(i,j) = (0.d0,0.d0)\\
  ~~ end do\\
  end do \- \\
  {\it c} \> \+ {\it Example: diagonal entries normalized to Y\_l as in SUGRA}  \\
  slmi\_lr(1,1) = (1.d-4,0.d0) \> {\it ! A\_e}\\
  slmi\_lr(2,2) = (1.0d-2,0.d0) \> {\it ! A\_mu}\\
  slmi\_lr(3,3) = (1.0d-1,0.d0) \> {\it ! A\_tau}\\
  slmi\_lr(2,3) = (2.d-2,1.d-2) \> {\it ! example, non-vanishing LR 23 entry} \- \\
  {\it c} \> \+ {\it Calculate physical masses and mixing angles}\\
  call init\_slepton\_sector(sll,slr,slmi\_l,slmi\_r,slmi\_lr,slmi\_lrp \- \\
  ~~~~~~\$~~~~ ,ierr)\\
  \>     if (ierr.ne.0) stop 'negative tree level slepton mass$^2$?' \\
  \\
  {\it c} \> \+ {\it Squark diagonal soft breaking parameters}\\
  amsq(1) = 500.d0          \> {\it ! left squark mass, 1st generation}\\
  amsq(2) = 450.d0          \> {\it ! left squark mass, 2nd generation}\\
  amsq(3) = 400.d0          \> {\it ! left squark mass, 3rd generation}\\
  amsd(1) = 550.d0          \> {\it ! right down squark mass}\\
  amsd(2) = 550.d0          \> {\it ! right strange squark mass}\\
  amsd(3) = 300.d0          \> {\it ! right sbottom mass}\\
  amsu(1) = 450.d0          \> {\it ! right up squark mass}\\
  amsu(2) = 450.d0          \> {\it ! right charm squark mass}\\
  amsu(3) = 200.d0          \> {\it ! right stop mass} \- \\
  {\it c} \>  {\it Squark LL and RR mass insertions (hermitian matrices)}\\
  {\it c} \> \+ {\it sqmi\_l(1),sqmi\_l(2), sqmi\_l(3) are 12,23,31 entry respectively, etc.}\\
  do i=1,3\\
  ~~   sqmi\_l(i) = (0.d0,0.d0) \> {\it ! squark LL mass insertion}\\
  ~~   sumi\_r(i) = (0.d0,0.d0) \> {\it ! up-squark RR mass insertion}\\
  ~~   sdmi\_r(i) = (0.d0,0.d0) \> {\it ! down-squark RR mass insertion}\\
  end do \\
  sqmi\_l(2) = (-1.d-2,1.d-2) \> {\it ! example, non-vanishing LL 23 entry} \- \\
  {\it c} \> {\it Squark holomorphic LR mass insertions, non-hermitian in general}\\
  {\it c} \> \+ {\it All entries dimensionless (normalized to masses)}\\
  do i=1,3\\
  ~~   do j=1,3 \- \\
  {\it c} \> \+ {\it holomorphic LR mixing terms}\\
  ~~~~     sumi\_lr(i,j) = (0.d0,0.d0) \> {\it ! up-squark }\\
  ~~~~     sdmi\_lr(i,j) = (0.d0,0.d0) \> {\it ! down-squark } \- \\
  {\it c} \> \+ {\it non-holomorphic LR mixing terms}\\
  ~~~~     sumi\_lrp(i,j) = (0.d0,0.d0) \> {\it ! up-squark}\\
  ~~~~     sdmi\_lrp(i,j) = (0.d0,0.d0) \> {\it ! down-squark}\\
  ~~  end do\\
  end do \- \\
  {\it c} \> \+ {\it Example: diagonal entries normalized to Y\_d,Y\_u as in SUGRA}  \\
  sumi\_lr(1,1) = dcmplx(1.d-5,0.d0)\\
  sumi\_lr(2,2) = dcmplx(4.d-3,0.d0)\\
  sumi\_lr(3,3) = dcmplx(1.d0,0.d0)\\
  sdmi\_lr(1,1) = dcmplx(-1.d-3,0.d0)\\
  sdmi\_lr(2,2) = dcmplx(-2.d-2,0.d0)\\
  sdmi\_lr(3,3) = dcmplx(-8.d-1,0.d0)\\
    sdmi\_lr(2,3) = (1.d-2,-1.d-2) \> {\it ! example, non-vanishing down LR 23 entry} \- \\
  {\it c} \> \+ {\it Calculate physical masses and mixing angles}\\
  call init\_squark\_sector(amsq,amsu,amsd,sqmi\_l,sumi\_r,sdmi\_r, \- \\
  ~~~~~~\$~~~~    sumi\_lr,sdmi\_lr,sumi\_lrp,sdmi\_lrp,ierr)\\
  \>    if (ierr.ne.0) stop 'negative tree level squark mass$^2$?' \\
  \\
  {\it c} \> \+ {\it reset status of physical Higgs mass after parameter changes}\\
  call reset\_phys\_data \- \\
\\

  {\it c} \> \+ {\it  Neutral CP-even Higgs masses with the 2-loop approximate formula}\\
  call mhcorr\_app2(ierr)\-\\
  \>   if (ierr.ne.0) stop 'negative CP-even Higgs mass$^2$?'  \\
  \\
  {\it c} \> \+ {\it !!! End of input section !!!}\\
\\
  call set\_resummation\_level(ilev,ierr)  \\
  if (ierr.ne.0) write(*,*)ierr,'Error in chiral corrections
  resummation!'\\
  call susy\_flavor \> {\it ! main routine calculating physical  observables} \\
  call sflav\_output(ilev,ierr) \> {\it ! output written to susy\_flavor.out} \\
  end  

\end{tabbing}
}
\end{small}

\newpage

\section{Example of  \code{} input file}
\label{app:infile}

By default, the driver program {\tt susy\_flavor\_file.f} reads input
parameters from the file {\tt susy\_flavor.in}.  Starting from v2.5,
\code{} should be able to directly read most of output files defining
MSSM Lagrangian parameters produced by other public SUSY generators,
simply after renaming them to {\tt susy\_flavor.in}.  However, as
there are already many of such programs and they do not always
uniformly follow SLHA2 standards, some incompatibilities may
eventually occur.  In such case, please send a message to program
maintainer, so the problem could be removed in next versions of
\code{} library.

Below we provide an example input file defining a set of parameters
equivalent to those given in the driver file presented in
Appendix~\ref{app:code}.
\begin{small}
{\tt
\begin{tabbing}
~~~~\=~~~~~\=~~~~~~~~~~~~~~~~~~~~~~~~~~~~\=~~~~~\=\\[-6mm]
\# Example input of SUSY\_FLAVOR in Les Houches-like format\\
\#\\
Block MODSEL\>	\>	 \>  \# Select model\\
\>  1 \>   0		 \>  \# General MSSM\\
\>  3\>	 0		 \>  \# MSSM particle content\\
\>  4\>	 0		 \>  \# R-parity conserving MSSM\\
\>  5\>	 2		 \>  \# CP violated\\
\>  6\>	 3		 \>  \# Lepton and quark flavor violated\\
Block SOFTINP\>	\>	 \>  \# Choose convention for the soft terms\\
\#\\
\# Block SOFTINP is optional - standard SLHA2 used if it is missing,\\
\# i.e. convention=1, input\_type=2, ilev=2. Otherwise:\\
\#\\
\# convention = 1(2): input parameters in SLHA2(hep-ph/9511250) conventions \\
\# input\_type = 1:\\
\#   sfermion off-diagonal terms given as dimensionless mass insertions\\
\#   LR diagonal terms given as dimensionless parameters\\
\# input\_type = 2:\\
\#   sfermion soft terms given as absolute values (default)\\
\# ilev = 0\\
\#   no resummation of chirally enhanced corrections\\
\# ilev = 1\\
\#   analytical resummation of chirally enhanced corrections \\
\#   in the limit v1,v2 << M\_SUSY\\
\# ilev = 2 (default)\\
\#   numerical iterative resummation of chirally enhanced corrections \\
\# See comment in Blocks MSXIN2, TXIN below\\
\>  1\>	 1		 \>  \# iconv (conventions, SLHA2 or hep-ph/9511250)\\
\>  2  \>  2		 \>  \# input\_type (dimension of soft mass entries)\\
\>  3\>	 2		 \>  \# ilev (level of chiral corrections resummation) \\
Block SMINPUTS	\>	\> \>  \# Standard Model inputs\\
\>  1\>	1.279340000e+02	 \>  \# alpha$^(-1)$ SM MSbar(MZ)\\
\>  3 \>  1.172000000e-01	 \>  \# alpha\_s(MZ) SM MSbar\\
\>  4 \>  9.118760000e+01	 \>  \# MZ(pole)\\
\>  5\>	4.180000000e+00	 \>  \# mb(mb) SM MSbar\\
\>  6 \>  1.735000000e+02	 \>  \# mtop(pole)\\
\>  7\>	 1.77684000000e+00	 \>  \# mtau(pole)\\
\>  11\> 5.10998900000e-04	 \>  \# me(pole)\\
\>  13\>	1.056580000e-01  \>  \# mmu(pole)\\
\>  21\>	4.700000000e-03	 \>  \# md(2 GeV) MSbar\\
\>  22\>	2.100000000e-03	 \>  \# mu(2 GeV) MSbar\\
\>  23\>	9.340000000e-02	 \>  \# ms(2 GeV) MSbar\\
\>  24\>	1.279000000e+00	 \>  \# mc(mc) MSbar\\
\>  30\>  8.039800000e+01  \>  \# MW (pole), not a standard SLHA2 entry !!!\\
\>  31\>  2.31160000000e-01  \>  \# sW$^2$ (MSBar), not a standard SLHA2 entry !!!\\
Block VCKMIN\>\>		 \>  \# CKM matrix\\
\>  1 \>  2.258000000e-01	 \>  \# lambda\\
\>  2 \>  8.080000000e-01	 \>  \# A\\
\>  3 \>  1.770000000e-01	 \>  \# rho bar\\
\>  4 \>  3.600000000e-01	 \>  \# eta bar\\
Block EXTPAR   \>  \>        \>  \# non-minimal input parameters, real part\\
\>  0 \> -1.000000000e+00	 \>  \# input at EW scale only, cannot be modified!!!\\
\>  1 \>  2.000000000e+02  \>  \# Re(m1), U(1) gaugino mass\\
\>  2 \>  3.000000000e+02  \>  \# Re(m2), SU(2) gaugino mass\\
\>  3 \>  6.000000000e+02  \>  \# m3, SU(3) gaugino mass\\
\>  23\>	2.000000000e+02	 \>  \# Re(mu) \\
\>  25\>	4.000000000e+00	 \>  \# tan(beta)\\
\>  26\>	2.000000000e+02	 \>  \# MA\\
Block IMEXTPAR  \>     \>    \>  \# non-minimal input parameters, imaginary part\\
\>  1 \> 0.000000000e+00  \>  \# Im(m1), U(1) gaugino mass\\
\>  2 \>  0.000000000e+00  \>  \# Im(m2), SU(2) gaugino mass\\
\>  23\>	1.000000000e+02  \>  \# Im(mu)\\
\# if abs(m1) = 0 SUSY\_FLAVOR uses m1=5/3 $s_W^2/c_W^2$ m2\\
\#   \\
\# Soft sfermion mass matrices\\
\#\\
\# Off-diagonal entries may be given as absolute entries or as\\
\# dimensionless mass insertions - then real off-diagonal entries of\\
\# SLHA2 blocks are calculated by SUSY\_FLAVOR as\\
\# M$^2$(I,J) = (mass insertion)(I,J) sqrt(M$^2$(I,I) M$^2$(J,J))\\
\# (see comments at the top of subroutine sflav\_input) \\
\#\\
\# Below we give an example of dimensionful off-diagonal entries\\
\#\\
Block MSL2IN    \>   \>      \>  \# left soft slepton mass matrix, real part\\
\> 1  1 \> 9.000000000e+04	 \>  \# Left slepton diagonal mass$^2$, 1st generation\\
\> 2  2 \> 9.000000000e+04	 \>  \# Left slepton diagonal mass$^2$, 2nd generation\\
\> 3  3 \> 9.000000000e+04	 \>  \# Left slepton diagonal mass$^2$, 3rd generation\\
\> 1  2 \> 1.800000000e-02   \>  \# Left slepton mass insertion 12\\
\> 2  3\>  0.000000000e+00   \>  \# Left slepton mass insertion 23\\
\> 1  3\>  0.000000000e+00   \>  \# Left slepton mass insertion 13\\
Block IMMSL2IN    \>   \>    \>  \# Left soft slepton mass matrix, imaginary part\\
\> 1  2\>  2.700000000e+03   \>  \# Left slepton mass insertion 12\\
\> 2  3\>  0.000000000e+00   \>  \# Left slepton mass insertion 23\\
\> 1  3 \> 0.000000000e+00   \>  \# Left slepton mass insertion 13\\
Block MSE2IN     \>    \>    \>  \# right soft slepton mass matrix, real part\\
\> 1  1\>  9.000000000e+04	 \>  \# Right selectron diagonal mass$^2$\\
\> 2  2\>  9.000000000e+04	 \>  \# Right smuon diagonal mass$^2$\\
\> 3  3 \> 9.000000000e+04	 \>  \# Right stau diagonal mass$^2$\\
\> 1  2 \> 0.000000000e+00   \>  \# right slepton mass insertion 12\\
\> 2  3\>  0.000000000e+00   \>  \# right slepton mass insertion 23\\
\> 1  3\>  0.000000000e+00   \>  \# right slepton mass insertion 13\\
Block IMMSE2IN   \>   \>     \>  \# right soft slepton mass matrix, imaginary part\\
\> 1  2\>  0.000000000e+00	 \>  \# right slepton mass insertion 12\\
\> 2  3 \> 0.000000000e+00   \>  \# right slepton mass insertion 23\\
\> 1  3\>  0.000000000e+00   \>  \# right slepton mass insertion 13\\
Block MSQ2IN   \>     \>     \>  \# Left soft squark mass matrix, real part\\
\> 1  1 \> 2.500000000e+05	 \>  \# Left squark diagonal mass$^2$, 1st generation\\
\> 2  2\>  2.025000000e+05	 \>  \# Left squark diagonal mass$^2$, 2nd generation\\
\> 3  3 \> 1.600000000e+05	 \>  \# Left squark diagonal mass$^2$, 3rd generation\\
\> 1  2 \> 0.000000000e+00   \>  \# Left squark mass insertion 12\\
\> 2  3 \> -1.800000000e+03   \>  \# Left squark mass insertion 23\\
\> 1  3\>  0.000000000e+00   \>  \# Left squark mass insertion 13\\
Block IMMSQ2IN   \>    \>    \>  \# Left soft squark mass matrix, imaginary part\\
\> 1  2 \> 0.000000000e+00   \>  \# Left squark mass insertion 12\\
\> 2  3\>  1.800000000e+03   \>  \# Left squark mass insertion 23\\
\> 1  3\>  0.000000000e+00   \>  \# Left squark mass insertion 13\\
Block MSU2IN    \>   \>      \>  \# Right soft up-squark mass matrix, real part\\
\> 1  1\>  2.025000000e+05	 \>  \# Right u-squark diagonal mass$^2$\\
\> 2  2\>  2.025000000e+05	 \>  \# Right c-squark diagonal mass$^2$\\
\> 3  3\>  4.000000000e+04	 \>  \# Right stop diagonal mass$^2$\\
\> 1  2 \> 0.000000000e+00   \>  \# Right up-squark mass insertion 12\\
\> 2  3\>  0.000000000e+00   \>  \# Right up-squark mass insertion 23\\
\> 1  3 \> 0.000000000e+00   \>  \# Right up-squark mass insertion 13\\
Block IMMSU2IN     \>   \>   \>  \# Right soft up-squark mass matrix, imaginary part\\
\> 1  2 \> 0.000000000e+00   \>  \# Right up-squark mass insertion 12\\
\> 2  3 \> 0.000000000e+00   \>  \# Right up-squark mass insertion 23\\
\> 1  3 \> 0.000000000e+00   \>  \# Right up-squark mass insertion 13\\
Block MSD2IN \>  \>          \>  \# Right soft down-squark mass matrix, real part\\
\> 1  1 \> 3.025000000e+05	 \>  \# Right d-squark diagonal mass$^2$\\
\> 2  2 \> 3.025000000e+05	 \>  \# Right s-squark diagonal mass$^2$\\
\> 3  3 \> 9.000000000e+04	 \>  \# Right sbottom diagonal mass$^2$\\
\> 1  2 \> 0.000000000e+00   \>  \# Right down-squark mass insertion 12\\
\> 2  3 \> 0.000000000e+00   \>  \# Right down-squark mass insertion 23\\
\> 1  3 \> 0.000000000e+00   \>  \# Right down-squark mass insertion 13\\
Block IMMSD2IN \>    \>      \>  \# Right soft down-squark mass matrix, imaginary part\\
\> 1  2 \> 0.000000000e+00   \>  \# Right down-squark mass insertion 12\\
\> 2  3 \> 0.000000000e+00   \>  \# Right down-squark mass insertion 23\\
\> 1  3 \> 0.000000000e+00   \>  \# Right down-squark mass insertion 13\\
\#\\
\# Soft sfermion trilinear mixing matrices\\
\#\\
\# LR mixing parameters can be given as absolute entries or as\\
\# dimensionless diagonal A-terms and dimensionless off-diagonal mass\\
\# insertions - see comments at the top of subroutine sflav\_input\\
\#\\
\# In the second case the dimensionful entries of LR blocks \\
\# are calculated by SUSY\_FLAVOR as\\
\# TL(I,J) = AL(I,J) (ML$^2$(I,I)*ME$^2$(J,J))**(1/4)\\
\# TU(I,J) = AU(I,J) (MQ$^2$(I,I)*MU$^2$(J,J))**(1/4)\\
\# TD(I,J) = AD(I,J) (MQ$^2$(I,I)*MD$^2$(J,J))**(1/4)\\
\#\\
\# Below we give an example of dimensionful ``A terms''.\\
\#\\
Block TEIN  \>   \>          \>  \# slepton trilinear mixing, real part\\
\> 1  1 \> 3.000000000e-02	 \>  \# Diagonal AL term, 1st generation\\
\> 2  2 \> 3.000000000e-00   \>  \# Diagonal AL term, 2nd generation\\
\> 3  3 \> 3.000000000e+01   \>  \# Diagonal AL term, 3rd generation\\
\> 1  2 \> 0.000000000e+00   \>  \# Slepton LR mass insertion 12\\
\> 2  1 \> 0.000000000e+00   \>  \# Slepton LR mass insertion 21\\
\> 2  3 \> 2.000000000e-02   \>  \# Slepton LR mass insertion 23\\
\> 3  2 \> 0.000000000e+00   \>  \# Slepton LR mass insertion 32\\
\> 1  3 \> 0.000000000e+00   \>  \# Slepton LR mass insertion 13\\
\> 3  1\>  0.000000000e+00   \>  \# Slepton LR mass insertion 31\\
Block IMTEIN    \>  \>       \>  \# slepton trilinear mixing, imaginary part\\
\> 1  1\>  0.000000000e+00	 \>  \# Diagonal AL term, 1st generation\\
\> 2  2\>  0.000000000e+00   \>  \# Diagonal AL term, 2nd generation\\
\> 3  3\>  0.000000000e+00   \>  \# Diagonal AL term, 3rd generation\\
\> 1  2\>  0.000000000e+00   \>  \# Slepton LR mass insertion 12\\
\> 2  1\>  0.000000000e+00   \>  \# Slepton LR mass insertion 21\\
\> 2  3\>  1.000000000e-02   \>  \# Slepton LR mass insertion 23\\
\> 3  2\>  0.000000000e+00   \>  \# Slepton LR mass insertion 32\\
\> 1  3 \> 0.000000000e+00   \>  \# Slepton LR mass insertion 13\\
\> 3  1 \> 0.000000000e+00   \>  \# Slepton LR mass insertion 31\\
Block TUIN    \>   \>        \>  \# up-squark trilinear mixing, real part\\
\> 1  1 \> 4.743000000e-03	 \>  \# Diagonal AU term, 1st generation\\
\> 2  2 \> 1.800000000e-00	 \>  \# Diagonal AU term, 2nd generation\\
\> 3  3 \> 2.828000000e+02	 \>  \# Diagonal AU term, 3rd generation\\
\> 1  2\>  0.000000000e+00   \>  \# Up-squark LR mass insertion 12\\
\> 2  1\>  0.000000000e+00   \>  \# Up-squark LR mass insertion 21\\
\> 2  3 \> 0.000000000e+00   \>  \# Up-squark LR mass insertion 23\\
\> 3  2 \> 0.000000000e+00   \>  \# Up-squark LR mass insertion 32\\
\> 1  3 \> 0.000000000e+00   \>  \# Up-squark LR mass insertion 13\\
\> 3  1\>  0.000000000e+00   \>  \# Up-squark LR mass insertion 31\\
Block IMTUIN    \>   \>      \>  \# up-squark trilinear mixing, imaginary part\\
\> 1  1 \> 0.000000000e+00	 \>  \# Diagonal AU term, 1st generation\\
\> 2  2 \> 0.000000000e+00	 \>  \# Diagonal AU term, 2nd generation\\
\> 3  3 \> 0.000000000e+00	 \>  \# Diagonal AU term, 3rd generation\\
\> 1  2 \> 0.000000000e+00   \>  \# Up-squark LR mass insertion 12\\
\> 2  1 \> 0.000000000e+00   \>  \# Up-squark LR mass insertion 21\\
\> 2  3 \> 0.000000000e+00   \>  \# Up-squark LR mass insertion 23\\
\> 3  2\>  0.000000000e+00   \>  \# Up-squark LR mass insertion 32\\
\> 1  3 \> 0.000000000e+00   \>  \# Up-squark LR mass insertion 13\\
\> 3  1 \> 0.000000000e+00   \>  \# Up-squark LR mass insertion 31\\
Block TDIN   \>  \>          \>  \# down-squark trilinear mixing, real part\\
\> 1  1\> -5.244000000e-02	 \>  \# Diagonal AD term, 1st generation\\
\> 2  2\> -9.950000000e-01	 \>  \# Diagonal AD term, 2nd generation\\
\> 3  3\> -2.771000000e+01	 \>  \# Diagonal AD term, 3rd generation\\
\> 1  2\>  0.000000000e+00   \>  \# Down-squark LR mass insertion 12\\
\> 2  1\>  0.000000000e+00   \>  \# Down-squark LR mass insertion 21\\
\> 2  3\>  1.000000000e-02   \>  \# Down-squark LR mass insertion 23\\
\> 3  2 \> 0.000000000e+00   \>  \# Down-squark LR mass insertion 32\\
\> 1  3 \> 0.000000000e+00   \>  \# Down-squark LR mass insertion 13\\
\> 3  1\>  0.000000000e+00   \>  \# Down-squark LR mass insertion 31\\
Block IMTDIN  \>   \>        \>  \# down-squark trilinear mixing, imaginary part\\
\> 1  1 \> 0.000000000e+00	 \>  \# Diagonal AD term, 1st generation\\
\> 2  2 \> 0.000000000e+00	 \>  \# Diagonal AD term, 2nd generation\\
\> 3  3 \> 0.000000000e+00	 \>  \# Diagonal AD term, 3rd generation\\
\> 1  2\>  0.000000000e+00   \>  \# Down-squark LR mass insertion 12\\
\> 2  1\>  0.000000000e+00   \>  \# Down-squark LR mass insertion 21\\
\> 2  3\>  -3.674000000e+00   \>  \# Down-squark LR mass insertion 23\\
\> 3  2 \> 0.000000000e+00   \>  \# Down-squark LR mass insertion 32\\
\> 1  3 \> 0.000000000e+00   \>  \# Down-squark LR mass insertion 13\\
\> 3  1 \> 0.000000000e+00   \>  \# Down-squark LR mass insertion 31\\
\#\\
\# ``Non-holomorphic'' soft sfermion trilinear mixing matrices (optional)\\
\# Such couplings are not SLHA2-standard and set to 0 if not explicitly\\
\# defined in the input file\\
\#\\
\# again LR mixing parameters can be given as absolute entries or as\\
\# dimensionless diagonal A-terms and dimensionless off-diagonal mass insertions\\
\#\\
Block TEINH  \>   \>          \>  \# slepton trilinear mixing, real part\\
\> 1  1 \> 0.000000000e-00	 \>  \# Diagonal ALNH term, 1st generation\\
\> 2  2 \> 0.000000000e-00   \>  \# Diagonal ALNH term, 2nd generation\\
\> 3  3 \> 0.000000000e-00   \>  \# Diagonal ALNH term, 3rd generation\\
\> 1  2 \> 0.000000000e+00   \>  \# Slepton LRNH mass insertion 12\\
\> 2  1 \> 0.000000000e+00   \>  \# Slepton LRNH mass insertion 21\\
\> 2  3 \> 0.000000000e-00   \>  \# Slepton LRNH mass insertion 23\\
\> 3  2 \> 0.000000000e+00   \>  \# Slepton LRNH mass insertion 32\\
\> 1  3 \> 0.000000000e+00   \>  \# Slepton LRNH mass insertion 13\\
\> 3  1\>  0.000000000e+00   \>  \# Slepton LRNH mass insertion 31\\
Block IMTEINH    \>  \>       \>  \# slepton trilinear mixing, imaginary part\\
\> 1  1\>  0.000000000e+00	 \>  \# Diagonal ALNH term, 1st generation\\
\> 2  2\>  0.000000000e+00   \>  \# Diagonal ALNH term, 2nd generation\\
\> 3  3\>  0.000000000e+00   \>  \# Diagonal ALNH term, 3rd generation\\
\> 1  2\>  0.000000000e+00   \>  \# Slepton LRNH mass insertion 12\\
\> 2  1\>  0.000000000e+00   \>  \# Slepton LRNH mass insertion 21\\
\> 2  3\>  0.000000000e-00   \>  \# Slepton LRNH mass insertion 23\\
\> 3  2\>  0.000000000e+00   \>  \# Slepton LRNH mass insertion 32\\
\> 1  3 \> 0.000000000e+00   \>  \# Slepton LRNH mass insertion 13\\
\> 3  1 \> 0.000000000e+00   \>  \# Slepton LRNH mass insertion 31\\
Block TUINH    \>   \>        \>  \# up-squark trilinear mixing, real part\\
\> 1  1 \> 0.000000000e-00	 \>  \# Diagonal AUNH term, 1st generation\\
\> 2  2 \> 0.000000000e-00	 \>  \# Diagonal AUNH term, 2nd generation\\
\> 3  3 \> 0.000000000e+00	 \>  \# Diagonal AUNH term, 3rd generation\\
\> 1  2\>  0.000000000e+00   \>  \# Up-squark LRNH mass insertion 12\\
\> 2  1\>  0.000000000e+00   \>  \# Up-squark LRNH mass insertion 21\\
\> 2  3 \> 0.000000000e-00   \>  \# Up-squark LRNH mass insertion 23\\
\> 3  2 \> 0.000000000e+00   \>  \# Up-squark LRNH mass insertion 32\\
\> 1  3 \> 0.000000000e+00   \>  \# Up-squark LRNH mass insertion 13\\
\> 3  1\>  0.000000000e+00   \>  \# Up-squark LRNH mass insertion 31\\
Block IMTUINH    \>   \>      \>  \# up-squark trilinear mixing, imaginary part\\
\> 1  1 \> 0.000000000e+00	 \>  \# Diagonal AUNH term, 1st generation\\
\> 2  2 \> 0.000000000e+00	 \>  \# Diagonal AUNH term, 2nd generation\\
\> 3  3 \> 0.000000000e+00	 \>  \# Diagonal AUNH term, 3rd generation\\
\> 1  2 \> 0.000000000e+00   \>  \# Up-squark LRNH mass insertion 12\\
\> 2  1 \> 0.000000000e+00   \>  \# Up-squark LRNH mass insertion 21\\
\> 2  3 \> 0.000000000e-00   \>  \# Up-squark LRNH mass insertion 23\\
\> 3  2\>  0.000000000e+00   \>  \# Up-squark LRNH mass insertion 32\\
\> 1  3 \> 0.000000000e+00   \>  \# Up-squark LRNH mass insertion 13\\
\> 3  1 \> 0.000000000e+00   \>  \# Up-squark LRNH mass insertion 31\\
Block TDINH   \>  \>          \>  \# down-squark trilinear mixing, real part\\
\> 1  1\>  0.000000000e-00	 \>  \# Diagonal ADNH term, 1st generation\\
\> 2  2\>  0.000000000e-00	 \>  \# Diagonal ADNH term, 2nd generation\\
\> 3  3\>  0.000000000e-00	 \>  \# Diagonal ADNH term, 3rd generation\\
\> 1  2\>  0.000000000e+00   \>  \# Down-squark LRNH mass insertion 12\\
\> 2  1\>  0.000000000e+00   \>  \# Down-squark LRNH mass insertion 21\\
\> 2  3\>  0.000000000e+00   \>  \# Down-squark LRNH mass insertion 23\\
\> 3  2 \> 0.000000000e+00   \>  \# Down-squark LRNH mass insertion 32\\
\> 1  3 \> 0.000000000e+00   \>  \# Down-squark LRNH mass insertion 13\\
\> 3  1\>  0.000000000e+00   \>  \# Down-squark LRNH mass insertion 31\\
Block IMTDINH  \>   \>        \>  \# down-squark trilinear mixing, imaginary part\\
\> 1  1 \> 0.000000000e+00	 \>  \# Diagonal ADNH term, 1st generation\\
\> 2  2 \> 0.000000000e+00	 \>  \# Diagonal ADNH term, 2nd generation\\
\> 3  3 \> 0.000000000e+00	 \>  \# Diagonal ADNH term, 3rd generation\\
\> 1  2\>  0.000000000e+00   \>  \# Down-squark LRNH mass insertion 12\\
\> 2  1\>  0.000000000e+00   \>  \# Down-squark LRNH mass insertion 21\\
\> 2  3\>  0.000000000e+00   \>  \# Down-squark LRNH mass insertion 23\\
\> 3  2 \> 0.000000000e+00   \>  \# Down-squark LRNH mass insertion 32\\
\> 1  3 \> 0.000000000e+00   \>  \# Down-squark LRNH mass insertion 13\\
\> 3  1 \> 0.000000000e+00   \>  \# Down-squark LRNH mass insertion 31\\
Block SFLAV\_HADRON \>   \>        \>  \# hadronic and QCD-related input\\
\> 1    \>   0.1561e0 \>	\# f\_K\\
\> 2   \>    0.2e0     \>	\# f\_D\\
\> 3    \>   0.193e0   \>	\# f\_B\_d\\
\> 4  \>  0.232e0\>	\# f\_B\_s\\
\> 5  \>  0.724e0   \>	\# B\_K for SM contribution to KKbar\\
\> 6  \>  1.86e0   \>	\# eta\_cc in KK mixing (SM)\\
\> 7  \>  0.496e0   \>	\# eta\_ct in KK mixing (SM)\\
\> 8  \>  0.577e0   \>	\# eta\_tt in KK mixing (SM)\\
\> 9  \>  2.e0   \>	\# scale for B\_K (non-SM)    \\
\> 10  \> 0.61e0  \>	\# B\_K for VLL (non-SM)\\
\> 11  \> 0.76e0  \>	\# B\_K for SLL1\\
\> 12  \> 0.51e0  \>	\# B\_K for SLL2\\
\> 13  \> 0.96e0  \>	\# B\_K for LR1 \\
\> 14  \> 1.30e0  \>	\# B\_K for LR2 \\
\> 15  \> 1.e0    \>	\# B\_D for SM contribution \\
\> 16  \> 2.e0    \>	\# scale for B\_D (non-SM)\\
\> 17  \> 1.e0    \>	\# B\_D for VLL\\
\> 18  \> 1.e0    \>	\# B\_D for SLL1\\
\> 19  \> 1.e0    \>	\# B\_D for SLL2\\
\> 20  \> 1.e0    \>	\# B\_D for LR1 \\
\> 21  \> 1.e0    \>	\# B\_D for LR2 \\
\> 22  \> 1.22e0  \>	\# B\_Bd for SM contribution \\
\> 23  \> 4.6e0   \>	\# scale for B\_B (non-SM, both Bd and Bs)\\
\> 24  \> 0.87e0  \>	\# B\_Bd for VLL (non-SM)\\
\> 25  \> 0.8e0   \>	\# B\_Bd for SLL1\\
\> 26  \> 0.71e0  \>	\# B\_Bd for SLL2\\
\> 27  \> 1.71e0  \>	\# B\_Bd for LR1 \\
\> 28  \> 1.16e0  \>	\# B\_Bd for LR2 \\
\> 29  \> 1.22e0  \>	\# B\_Bs for SM contribution \\
\> 30  \> 0.55e0  \>	\# eta\_b for BsBs (SM)\\
\> 31  \> 0.87e0  \>	\# B\_Bs for VLL (non-SM)\\
\> 32  \> 0.8e0   \>	\# B\_Bs for SLL1\\
\> 33  \> 0.71e0  \>	\# B\_Bs for SLL2\\
\> 34  \> 1.71e0  \>	\# B\_Bs for LR1 \\
\> 35  \> 1.16e0  \>	\# B\_Bs for LR2 \\
\> 36  \> 1.519e-12 \>	\# Bd lifetime (experimental value)\\
\> 37  \> 1.512e-12 \>	\# Bs lifetime (experimental value)\\
\> 38  \> 5.27958e0 \>	\# Bd mass (experimental value)\\
\> 39  \> 5.36677e0 \>	\# Bs mass (experimental value)\\
\> 40  \> 3.337e-13 \>	\# Delta Bd (experimental value)\\
\> 41  \> 1.17e-11 \>	\# Delta Bs (experimental value)\\
\> 42  \> 0.497614e0 \>	\# K0 mass (experimental value)\\
\> 43  \> 3.483e-15 \>	\# Delta mK (experimental value)\\
\> 44  \> 2.229e-3 \>	\# eps\_K (experimental value)\\
\> 45  \> 1.8645e0 \>	\# D0 mass (experimental value)\\
\> 46  \> 1.56e-14 \>	\# Delta mD (experimental value)\\
\> 47  \> 2.231e-10 \>	\# parameter kappa in K$^0$->pi$^0$vv calculations\\
\> 48  \> 5.173e-11 \>	\# parameter kappa in K$^+$->pi$^+$vv calculations\\
\> 49  \> 0.41e0  \>	\# parameter P\_c in K->pivv calculations\\
\> 50  \>   0.013e-10 \>	\# error of kappa0\\
\> 51  \>   0.024e-11 \>	\# error of kappa+\\
\> 52  \>   0.03e0  \>	\# error of P\_c\\
\> 53  \>   0.79e0  \>	\# neutron EDM\_d QCD coefficient\\
\> 54  \>  -0.2e0   \>	\# neutron EDM\_u QCD coefficient\\
\> 55  \>   0.59e0  \> 	\# neutron CDM\_d QCD coefficient\\
\> 56  \>   0.3e0   \>	\# neutron CDM\_u QCD coefficient\\
\> 57  \>   3.4e0   \>	\# neutron CDM\_g QCD coefficient\\
\> 58  \>   1.18e0  \>	\# neutron EDM chiral symmetry breaking scale\\
\> 59  \>   1.5e0   \>	\# pole c quark mass (in B->X\_s gamma and t->cH)\\
\> 60  \>   0.1872e0   \> \# Br(tau->evv)\\
\> 61 \>    5.27917e0   \>\# M\_B+\\
\> 62  \>   0.297e0   \>\# Br(B->D tau nu)/Br(B->D l nu) in SM\\
\> 63  \>   0.017e0  \> \# error of Br(B->D tau nu)/Br(B->D l nu) in SM\\
\> 64  \>   0.252e0   \>\# Br(B->D* tau nu)/Br(B->D* l nu) in SM\\
\> 65  \>   0.003e0  \> \# error of Br(B->D* tau nu)/Br(B->D* l nu) in SM\\

\end{tabbing}
}
\end{small}

\newpage

\section{Example of  \code{} output}
\label{app:output}

The parameters defined inside the driver program in
Appendix~\ref{app:code} and in the input file listed in
Appendix~\ref{app:infile} should produce almost identical output, up
to minor differences on distant decimal digits coming from finite
accuracy of numerical computations.

 We enclose content of the {\tt susy\_flavor.out} output file here, so
 that \code{} users can check that the program gives the same result
 on their own computers and compilers.
\begin{small} 
{\tt 
\begin{tabbing}
~~~~\=~~~~~~~~~\=~~~~~~~~~~~~~~~~~~~~~\=~~~~~\=\\[-6mm]
 \#\\
 \#   \> \>     ***************************\\
 \#  \>   \>    * SUSY\_FLAVOR 2.50 output *\\
 \#    \>  \>   ***************************\\
 \#\\
 BLOCK SFLAV\_CONTROL\\
 \>   1  \>  2           \>           \# resummation level of chiral corrections\\
   \> 2  \>  0         \>             \# error code (0 if all calculations were correct)\\
 BLOCK SFLAV\_MASS  \>\> \>     \# Mass Spectrum\\
 \#  \>   code   \>       mass      \>    \# particle\\
   \>    24  \>    8.039800000E+01 \>   \# W+\\
    \>     25 \>     9.700316978E+01\>    \# h (simple 2-loop approximation only)\\
    \>     35  \>    2.061912209E+02 \>   \# H (simple 2-loop approximation only)\\
   \>      36 \>     2.000000000E+02 \>   \# A\\
    \>     37 \>     2.155547225E+02 \>   \# H+\\
   \>      41 \>     5.109989000E-04 \>   \# e (pole)\\
   \>      42  \>    1.056580000E-01 \>   \# mu (pole)\\
   \>      43 \>     1.776840000E+00 \>   \# tau (pole)\\
   \>      44 \>     2.608286100E-03 \>   \# md(mt) (running)\\
   \>      45 \>     5.183274930E-02 \>   \# ms(mt) (running)\\
   \>      46 \>     2.744876788E+00 \>   \# mb(mt) (running)\\
   \>      47  \>    1.165404427E-03 \>   \# mu(mt) (running)\\
    \>     48 \>     6.081579020E-01 \>   \# mc(mt) (running)\\
    \>     49  \>    1.630910000E+02 \>   \# mt(mt) (running)\\
\>  1000021   \>   6.000000000E+02 \>   \# $\sim$g\\
\>  1000022  \>    1.609162276E+02 \>   \# $\sim$chi\_10\\
\>  1000023  \>    2.232344115E+02  \>  \# $\sim$chi\_20\\
\>  1000025  \>    2.283407379E+02 \>   \# $\sim$chi\_30\\
\>  1000035  \>    3.446204135E+02 \>   \# $\sim$chi\_40\\
\>  1000024  \>    1.879079878E+02 \>   \# $\sim$chi\_1+\\
 \> 1000037  \>    3.427487004E+02 \>   \# $\sim$chi\_2+\\
 \# sfermion mass eigenstates\\
  \>     101    \>   3.006758739E+02   \>  \# $\sim$d(1)\\
  \>     102   \>    4.038884306E+02  \>   \# $\sim$d(2)\\
  \>     103   \>    4.536071034E+02  \>   \# $\sim$d(3)\\
   \>    104   \>    5.030960409E+02   \>  \# $\sim$d(4)\\
   \>    105  \>     5.505085911E+02  \>   \# $\sim$d(5)\\
   \>    106   \>    5.505109533E+02  \>   \# $\sim$d(6)\\
   \>    111   \>    2.322291420E+02  \>   \# $\sim$u(1)\\
   \>    112   \>    4.406135873E+02  \>   \# $\sim$u(2)\\
  \>     113   \>    4.486873688E+02  \>   \# $\sim$u(3)\\
  \>     114   \>    4.487396867E+02  \>   \# $\sim$u(4)\\
  \>     115   \>    4.498531834E+02  \>   \# $\sim$u(5)\\
  \>     116   \>    4.974625553E+02  \>   \# $\sim$u(6)\\
  \>     121   \>    2.978728202E+02  \>   \# $\sim$l(1)\\
  \>     122   \>    3.017183129E+02   \>  \# $\sim$l(2)\\
  \>     123   \>    3.028111419E+02  \>   \# $\sim$l(3)\\
  \>     124   \>    3.028119935E+02  \>   \# $\sim$l(4)\\
  \>     125  \>     3.043692532E+02  \>   \# $\sim$l(5)\\
  \>     126   \>    3.085762706E+02  \>   \# $\sim$l(6)\\
  \>     131   \>    2.882497056E+02   \>  \# $\sim$nu(1)\\
   \>    132   \>    2.938268860E+02   \>  \# $\sim$nu(2)\\
  \>     133   \>    2.992956484E+02   \>  \# $\sim$nu(3)\\
 BLOCK SFLAV\_CHIRAL\_YUKAWA \> \> \>  \# Chiral corrections size to Yukawa couplings\\
  \>      1  \>    9.250781508E-03  \>  \# correction to Y\_e\\
  \>      2  \>    7.871358686E-03  \>  \# correction to Y\_mu\\
  \>      3  \>    7.355398855E-03 \>   \# correction to Y\_tau\\
  \>      4  \>    2.825581825E-02 \>   \# correction to Y\_d\\
  \>      5  \>    2.875084532E-02 \>   \# correction to Y\_s\\
  \>      6   \>   4.067136212E-02 \>   \# correction to Y\_b\\
  \>      7  \>    1.478999649E-02 \>   \# correction to Y\_u\\
   \>     8  \>    1.118390358E-02 \>   \# correction to Y\_c\\
   \>     9  \>    8.435040750E-03 \>   \# correction to Y\_t\\
 BLOCK SFLAV\_CHIRAL\_CKM\> \> \>   \# Chiral corrections size to CKM matrix\\
 \>   1   1  \>    3.660227820E-05  \>  \# correction to V\_11\\
  \>  1   2  \>    6.792764515E-04 \>   \# correction to V\_12\\
  \>  1   3   \>   3.087095532E-03 \>   \# correction to V\_13\\
  \>  2   1   \>   6.871751836E-04 \>   \# correction to V\_21\\
  \>  2   2   \>   3.415695240E-05 \>   \# correction to V\_22\\
  \>  2   3   \>   5.961443433E-03 \>   \# correction to V\_23\\
  \>  3   1   \>   5.998183639E-03 \>   \# correction to V\_31\\
 \>   3   2   \>   5.944557565E-03 \>   \# correction to V\_32\\
 \>   3   3   \>   7.838728907E-06\>    \# correction to V\_33\\
 BLOCK SFLAV\_DELTA\_F0 \> \> \>  \# Delta F = 0 processes\\
  \>      1  \>   -1.496831513E-25 \>   \# EDM\_e\\
  \>      2  \>   -3.083776497E-23 \>   \# EDM\_mu\\
  \>      3  \>   -5.176903910E-22 \>   \# EDM\_tau\\
  \>      4  \>    2.759938107E-25 \>   \# neutron EDM\\
  \>      5  \>    9.398319525E-15 \>   \# (g-2)\_e/2, SUSY contribution\\
   \>     6  \>    4.843853089E-10 \>   \# (g-2)\_mu/2, SUSY contribution\\
   \>     7  \>    1.458383883E-07 \>   \# (g-2)\_tau/2, SUSY contribution\\
 BLOCK SFLAV\_DELTA\_F1 \> \> \>  \# Delta F = 1 processes\\
   \>     1  \>    2.343751393E-08 \>   \# Br(mu-> e gamma)\\
   \>     2  \>    3.014685213E-20 \>   \# Br(tau-> e gamma)\\
   \>     3  \>    3.472210147E-09 \>   \# Br(tau-> mu gamma)\\
    \>    4  \>    2.797259621E-11 \>   \# Br(K0 -> pi0 nu nu)\\
    \>    5  \>    7.705350370E-11 \>   \# Br(K+ -> pi+ nu nu)\\
 \>       6   \>     8.768756807E-05   \>   \# BR(B -> tau nu)\\
  \>      7   \>     2.962481261E-01   \>   \# BR(B -> D tau nu)/BR(B -> D l nu)\\
 \>       8   \>     2.519503431E-01   \>  \# BR(B -> D* tau nu)/BR(B -> D* l nu)\\
   \>     9  \>    6.933649703E-04 \>   \# BR(B -> X\_s gamma)\\
   \>    10  \>    6.246862414E-12 \>   \# BR(t -> u h)\\
   \>    11  \>    1.945309468E-10  \>  \# BR(t -> c h)\\
   \>    12  \>    2.686141823E-15 \>   \# BR(B\_d -> e e)\\
   \>    13  \>    1.147486285E-10 \>   \# BR(B\_d -> mu mu)\\
    \>   14  \>    2.402190900E-08  \>  \# BR(B\_d -> tau tau)\\
   \>    15  \>    8.309954374E-22  \>  \# BR(B\_d -> mu e)\\
   \>    16  \>    6.024806941E-34 \>   \# BR(B\_d -> tau e)\\
   \>    17  \>    6.589103387E-24 \>   \# BR(B\_d -> tau mu)\\
   \>    18  \>    8.986707552E-14 \>   \# BR(B\_s -> e e)\\
   \>    19  \>    3.839108868E-09 \>   \# BR(B\_s -> mu mu)\\
    \>   20   \>   8.143373208E-07 \>   \# BR(B\_s -> tau tau)\\
    \>   21   \>   5.851883404E-20 \>   \# BR(B\_s -> mu e)\\
   \>    22   \>   2.300098073E-28  \>  \# BR(B\_s -> tau e)\\
   \>    23   \>   4.505919262E-23  \>  \# BR(B\_s -> tau mu)\\
 BLOCK SFLAV\_DELTA\_F2  \>\> \>  \# Delta F = 2 processes\\
  \>      1  \>    2.271797243E-03  \>  \# epsilon\_K\\
   \>     2  \>    2.324836393E-15 \>   \# Delta m\_K (GeV)\\
  \>      3  \>    1.792385187E-15 \>   \# Delta m\_D (GeV)\\
  \>      4  \>    3.520066391E-13 \>   \# Delta m\_Bd (GeV)\\
  \>      5  \>    1.195408141E-13 \>   \# Re(H\_eff\_Bd)\\
  \>      6  \>   -1.291787996E-13 \>   \# Im(H\_eff\_Bd)\\
  \>      7  \>    1.214313594E-11 \>   \# Delta m\_Bs (GeV)\\
   \>     8  \>    6.067828156E-12\>    \# Re(H\_eff\_Bs)\\
  \>      9  \>    2.130706402E-13 \>   \# Im(H\_eff\_Bs)\\

\end{tabbing}
}
\end{small}

\end{appendix}

\newpage

\noindent{\bf PROGRAM SUMMARY}\\

\noindent
\begin{small}
  {\em Manuscript Title:}~ \code{} v2.5: a computational tool for FCNC
  and CP-violating processes in the MSSM\\
  {\em Authors:}~ A.~Crivellin, J.~Rosiek, P.~Chankowski, A.~Dedes,
  S.~J\"ager,  P.~Tanedo\\
  {\em Program Title:}~ \code{} v2.5 \\
  {\em Journal Reference:}                                      \\
  {\em Catalogue identifier:}                                   \\
  {\em Licensing provisions:}~ None\\
  {\em Programming language:}~Fortran 77\\
  {\em Operating system:}~Any, tested on Linux\\
  {\em Keywords:}~Supersymmetry, $K$ physics, $B$ physics, rare
  decays, CP-violation   \\
  {\em PACS:}~12.60.Jv, 13.20.He\\
  {\em Classification:}~11.6 Phenomenological and Empirical Models and Theories\\
  {\em Nature of problem:}\\
  Predicting CP-violating observables, meson mixing parameters and
  branching ratios for set of rare processes in the general R-parity
  conserving MSSM.    \\
  {\em Solution method:}\\
  We use standard quantum theoretical methods to calculate Wilson
  coefficients in MSSM and at one loop including QCD corrections at
  higher orders when this is necessary and possible.  \\
  {\em Restrictions:}\\
  The results apply only to the case of MSSM with R-parity
  conservation.\\
  {\em Unusual features:}\\
  {\em Running time:}\\
  For single parameter set below 1s in {\tt double precision}
  on  a personal computer\\
  {\em References:}

\begin{refnummer}

\item J.~Rosiek, P.~Chankowski, A.~Dedes, S.~Jager and
  P.~Tanedo,
  Comput.\ Phys.\ Commun.\ {\bf 181} (2010) 2180 [arXiv:1003.4260
  [hep-ph]].  

\item M.~Misiak, S.~Pokorski and J.~Rosiek, {\sl ``Supersymmetry and
    the FCNC effects''} Adv.\ Ser.\ Direct.\ High Energy Phys.\ {\bf
    15} (1998) 795 [arXiv:hep-ph/9703442].

\item S.~Pokorski, J.~Rosiek and C.~A.~Savoy,
  Nucl.\ Phys.\ B {\bf 570} (2000) 81 [arXiv:hep-ph/9906206].

\item A.~Buras, P.~Chankowski, J.~Rosiek and L.~Slawianowska,
  Nucl.\ Phys.\ B {\bf 659} (2003) 3 [arXiv:hep-ph/0210145].

\item A.~Buras, T.~Ewerth, S.~Jager and J.~Rosiek,
  Nucl.\ Phys.\ B {\bf 714} (2005) 103 [arXiv:hep-ph/0408142].

\item A.~Dedes, J.~Rosiek and P.~Tanedo,
  Phys.\ Rev.\ D {\bf 79} (2009) 055 [arXiv:0812.4320 [hep-ph]].

\item A.~Crivellin, L.~Hofer and J.~Rosiek,
  JHEP {\bf 1107} (2011) 017 [arXiv:1103.4272
  [hep-ph]].  

\item A. Dedes, M.~Paraskevas, J. Rosiek, K. Suxho,
  K. Tamvakis [arXiv:1409.6546 [hep-ph]].

\end{refnummer}

\end{small}

\end{document}